\documentclass[11pt]{article}
\usepackage{url}
\usepackage{comment}
\usepackage{hyperref}
\usepackage{graphicx}
\usepackage{dcolumn}
\usepackage{bm}
\usepackage{amssymb}
\usepackage{amsmath, booktabs}
\usepackage{physics}
\usepackage{adjustbox}
\usepackage{parskip}
\usepackage{enumitem}
\usepackage{braket}
\usepackage{tabularx}
\usepackage[thinc]{esdiff}
\usepackage{physics}
\usepackage[thinlines]{easytable}

\usepackage{graphicx, subfigure,epstopdf, float}

\usepackage[utf8]{inputenc}
\usepackage[T1]{fontenc}
\usepackage{mathtools}

\usepackage[capitalise]{cleveref}
\usepackage{caption}
\usepackage{subcaption}

\usepackage{mdframed}
\usepackage{lipsum}

\hypersetup{
    colorlinks=true,
    linkcolor=blue,
    filecolor=magenta,      
    urlcolor=cyan,
    pdftitle={Overleaf Example},
    pdfpagemode=FullScreen,
}

\usepackage{amsmath,amsfonts,amssymb,amsthm}
\usepackage[margin=1in]{geometry}

\newcommand{\YY}{\ensuremath{\mathbf{Y}}}
\newcommand{\uu}{\ensuremath{\mathbf{u}}}

\newcommand{\yy}{\ensuremath{\mathbf{y}}}

\newcommand{\dt}[1]{\ensuremath{\frac{\dd#1}{\dd t}}}
\newcommand*{\QEDB}{\null\nobreak\hfill\ensuremath{\square}}
\newcommand{\uin}{\ensuremath{\norm{\mathbf{u}_{\mathrm{in}}}}}

\newcommand{\C}{{\mathbb{C}}}

\newcommand{\N}{{\mathbb{N}}}
\newcommand{\R}{{\mathbb{R}}}

\def\Re{\mathop{\rm Re}}
\def\C{{\mathbb C}}
\def\R{{\mathbb R}}
\def\N{{\mathbb N}}

\def\le{\leqslant}
\def\ge{\geqslant}

\newcommand{\Or}{\mathcal{O}}

\renewcommand{\d}{\mathrm{d}}

\renewcommand{\Re}{\mathop{\mathrm{Re}}}

\newcommand{\range}[1]{[{#1}]}
\newcommand{\rangez}[1]{[{#1}]_0}

\DeclareMathOperator{\poly}{poly}

\newtheorem{theorem}{Theorem}
\newtheorem{lemma}{Lemma}

\newtheorem{prob}{Problem}
\newtheorem{hypo}{Hypothesis}
\newtheorem{definition}{Definition}

\numberwithin{equation}{section}

\newcommand{\eq}[1]{(\ref{eq:#1})}
\renewcommand{\sec}[1]{\hyperref[sec:#1]{Section~\ref*{sec:#1}}}
\newcommand{\app}[1]{\hyperref[app:#1]{Appendix~\ref*{app:#1}}}
\newcommand{\thm}[1]{\hyperref[thm:#1]{Theorem~\ref*{thm:#1}}}
\newcommand{\lem}[1]{\hyperref[lem:#1]{Lemma~\ref*{lem:#1}}}
\newcommand{\cor}[1]{\hyperref[cor:#1]{Corollary~\ref*{cor:#1}}}
\newcommand{\prb}[1]{\hyperref[prb:#1]{Problem~\ref*{prb:#1}}}
\newcommand{\asmp}[1]{\hyperref[asmp:#1]{Assumption~\ref*{asmp:#1}}}
\newcommand{\hyp}[1]{\hyperref[hyp:#1]{Hypothesis~\ref*{hyp:#1}}}
\newcommand{\defn}[1]{\hyperref[defn:#1]{Definition~\ref*{defn:#1}}}

\newcommand\blfootnote[1]{%
  \begingroup
  \renewcommand\thefootnote{}\footnote{#1}%
  \addtocounter{footnote}{-1}%
  \endgroup
}

\begin{document}


\title{Two Quantum Algorithms for Nonlinear Reaction-Diffusion Equation using Chebyshev Approximation Method}

\author{Manish \ Kumar$^{1}$ \blfootnote{Email: manishkumar7@alum.iisc.ac.in} \\[2pt]
\small $^{1}$ Quantum Technology (IAP) \\
\small IISc Bengaluru
}

\date{\today}
\maketitle


\begin{abstract}
We present two new quantum algorithms for reaction-diffusion equations that employ the truncated Chebyshev polynomial approximation. This method is employed to numerically solve the ordinary differential equation emerging from the linearization of the associated nonlinear differential equation. In the first algorithm, we use the $\mathsf{matrix\ exponentiation}$ method (Patel et al., 2018), while in the second algorithm, we repurpose the $\mathsf{quantum\ spectral\ method}$ (Childs et al., 2020). Our main technical contribution is to derive the sufficient conditions for the diagonalization of the Carleman embedding matrix, which is indispensable for designing both quantum algorithms. We supplement this with an efficient iterative algorithm to diagonalize the Carleman matrix.\par

Our first algorithm has gate complexity of $\mathsf{O\big(d\cdot log(d)+T\cdot polylog({T}/{\varepsilon})\big)}$. Where $d$ is the size of the Carleman matrix, $T$ is the simulation time, and $\varepsilon$ is the approximation error. The second algorithm is polynomial in $\mathsf{log(d)}$, $T$, and $\mathsf{log(1/\varepsilon)}$ - the gate complexity scales as  $\mathsf{O\big(polylog(d)\cdot T\cdot polylog({T}/{\varepsilon})\big)}$. In terms of $\mathsf{T}$ and $\varepsilon$, this is comparable to the speedup gained by the current best known quantum algorithm for this problem, the $\mathsf{truncated\ Taylor\ series}$ method (Costa et.al., 2025). 
 
There are two shortcomings of our approach. First, we have not provided an upper bound, in terms of $d$, on the condition number of the Carleman matrix. Second, the success of the diagonalization is based on a conjecture that a specific trigonometric equation has no integral solution. However, we provide strategies to mitigate these shortcomings in most practical cases.
\end{abstract}
\setcounter{tocdepth}{2}
\tableofcontents

\newpage
\section{Introduction}
\label{sec:introduction}

\subsection{Quantum Algorithms for Linear Differential Equations}
Quantum Hamiltonian simulation has been considered a promising avenue for achieving a computational advantage over existing classical simulation methods \cite{Feynman_1982, seth96}. Basically, this is seeking a numerical solution for the Schrodinger equation, a linear ordinary differential equation (ODE) that governs the dynamics of closed quantum systems. Meanwhile, ODEs modeling physical systems are ubiquitous in science and engineering. Several classical numerical techniques have been developed over the past century \cite{Kress_1998}. Recently, numerous efforts have been made to design quantum algorithms for specific classical systems modeled by linear differential equations \cite{berry2017quantum, an2022theory}. Unlike classical algorithms, these quantum algorithms output a quantum state encoding the solution, rather than an explicit solution. \par

\subsection{Quantum algorithms for Non-linear Differential Equations}
Several natural systems, such as chemical reactions, fluid transport, and population dynamics, are well understood through the use of non-linear equations. For example, the Reaction-diffusion equation (RDE) for chemical systems and the Navier-Stokes equation (NSE) for fluid dynamics are widely studied nonlinear systems. As per the finite difference scheme, spatial discretization of the above two PDEs will produce non-linear ODEs \cite{Grossmann_Roos_Stynes_2016}. Recently, this observation has been used as an opportunity to explore quantum algorithms for such nonlinear systems by designing a quantum nonlinear ODE solver. However, the quantum algorithmic toolkit developed for linear ODEs can't be directly applied to non-linear ODEs, primarily because quantum dynamics is a linear theory. In a breakthrough result, \textbf{a special class} of non-linear ODEs, for a specific range of parameters, has been proven to be efficiently computable on a quantum computer \cite{Liu_Kolden_Krovi_Loureiro_Trivisa_Childs_2021}. These non-linear ODEs exhibit a special type of non-linearity called polynomial non-linearity. For example, the non-linear ODE
\begin{equation}
\diff{u}{t} = au + bu^{2} \qquad
\end{equation}
has quadratic non-linearity. An efficient quantum algorithm exists for this system. One of the key steps of the algorithm is to employ the Carleman linearization, a technique that embeds a non-linear ODE into an infinite system of linear ODEs. Now, the existing quantum algorithm for a linear system of ODEs can solve it by appropriately truncating the ODE system. \par
\subsection{Reaction-Diffusion Equation}
Reaction-Diffusion Equation (RDE) models various biological phenomena such as (i) leaf venation, (ii) tumor growth, and (iii) Turing patterns in tissues and organs \cite{Grindrod_1996}. RDE is a non-linear differential equation having the following one-dimensional form.

\begin{equation}
\pdv{u(x,t)}{t} = D\Delta u(x,t) + f(u(x,t)) \qquad 
\end{equation}
Here, $D\in \R$ is called the diffusion coefficient. The term $f(u(x,t))$ is referred to as the reaction term, which accounts for all local reactions.
Three practically useful cases are as follows.
\begin{enumerate}
    \item If $f(u) = au+bu^2$, it is called the $\mathsf{Fisher}$-$\mathsf{KPP}$ equation.
    \item If $f(u) = au+bu^3$, it is called the $\mathsf{Allen}$-$\mathsf{Kahn}$ equation. 
    \item If $f(u) = au^2+bu^3$, it is called the $\mathsf{Zeldovich}$ equation. 
\end{enumerate}
Throughout the paper, we use the $\mathsf{Fisher}$-$\mathsf{KPP}$ equation as a prototypical problem to design both our algorithms and demonstrate their effectiveness through numerical simulations. We also discuss how the techniques developed for the $\mathsf{Fisher}$-$\mathsf{KPP}$ equation can be extended to a related family of equations, the $\mathsf{Allen}$-$\mathsf{Kahn}$ equation being one example.
\section{Our Contributions \& Related Works}
Recently, two quantum algorithms have been proposed to solve it for a specific parameter regime- (i) forward Euler method \cite{Liu_Kolden_Krovi_Loureiro_Trivisa_Childs_2021} and (ii) Taylor series method \cite{costa2025further}. In the approximation theory literature, they are referred to as local approximation schemes, which closely approximate the function at the initial point. However, the approximation worsens as we move farther from the initial point. \par Meanwhile, the Chebyshev approximation connects variables globally \cite{Shen_Tang}. As a result, the error is nearly uniform throughout the domain. In fact, for specific functions, the Chebyshev series is known to converge faster than the Taylor series \footnote{A comprehensive analysis of the Chebyshev series can be found in \cite{powell01, Shen_Tang}.}. It motivated us to explore the Chebyshev approximation for solving the Carleman ODE problem, a crucial system of linear ODEs that arises in the context of solving certain reaction-diffusion equations. For further discussion, let us assume the system of ODEs is given by
\begin{equation}
    \diff{\yy}{t} = \textbf{A}\yy(t)
\end{equation}
where $\textbf{y}\in \R^{n}$, while $A\in \R^{n\times n}$.
Currently, there are two existing quantum algorithms for solving such a system of ODEs using the Chebyshev approximation method. They are (i) \textit{Matrix Exponentiation} \cite{patel_priyadarshini17} method and (ii) \textit{Quantum Spectral} \cite{Childs_Liu_2020} method. However, both algorithms require the matrix $A$ to be diagonalizable. The matrix $A$ is not a Hermitian matrix. In fact, it is not even a Normal matrix.\footnote{See this discussion on Math StackExchange \url{https://math.stackexchange.com/a/4870629/474528}} Thus, the method of its diagonalization is not very apparent, as we have for Normal matrices. \par 
Our first technical contribution is to provide the necessary and sufficient conditions for diagonalizing such matrices by exploiting their bi-diagonal/triangular block structure and special eigen-properties. In this pursuit, we have benefited from the recent study of \cite{wu2025quantum} on relating the No-resonance condition with Carleman matrices. The novelty we introduce is an iterative procedure for diagonalizing the matrix in the cases of the Fisher-KPP and related equations.\par
The rest of the paper is organized as follows. In Section \ref{sec: prob-setting}, we give a formal description of the problem and its variants. An overview of Carleman linearization and existing algorithms for the problem is provided in Section \ref{sec: overview} and Appendix \ref{apx: apx1}, respectively. Section \ref{sec:diagonalization} contains our analysis on the diagonalization of the Carleman matrix. The end-to-end design and analysis of both our algorithms are provided in Section \ref{sec: two-algo} and Section \ref{apx: simulation}.


\makebox[0pt][l]{%
\begin{minipage}{\textwidth}
\centering
    \includegraphics[width=.60\textwidth]{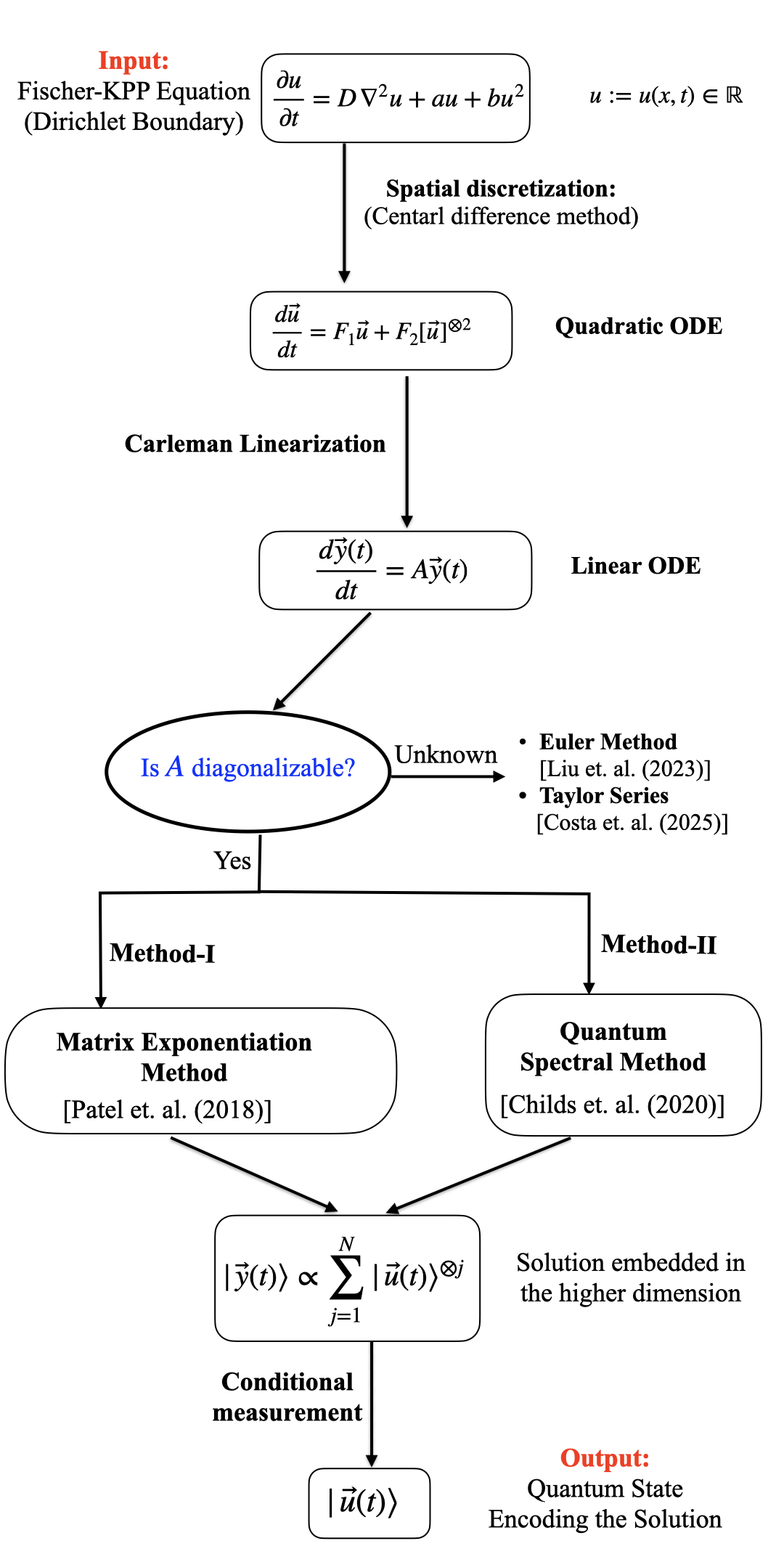}
 \captionof{figure}{Major steps in producing a quantum state encoding the solution of the $\mathsf{Fisher}$-$\mathsf{KPP}$ equation. We have investigated the possibility and advantage of using the Chebyshev approximation method to solve the emerging linear ODE -(i) $\mathsf{matrix\ exponentiation}$ method, and (ii) $\mathsf{quantum\ spectral}$ method. We overcome the main hurdle to adopting these two quantum ODE solvers — the diagonalization of the Carleman matrix ($A$) — by providing a necessary and sufficient condition for it. }
 \label{fig:summary_fig}
\end{minipage}
}
\section{Problem Setting}\label{sec: prob-setting}
In this section, we narrow our focus to a specific type of reaction-diffusion equation, known as the Fisher-Kolmogorov-Petrovskii-Piskunov equation. In a one-dimensional setting, it has the following form.

\begin{equation}
\frac{\partial u}{\partial t} = D \Delta u + au + bu^{2},
\label{eq:RDPDE}
\end{equation}

where, $u:=u(x,t)\in \R$. Let the domain of the problem be $x\in [0, 1]$ and  $t\in [0, T]$. It is a non-linear partial differential equation (PDE) if $b\neq 0$. With the finite difference method for spatial discretization, we obtain a quadratic ODE. In the next two subsections, we discuss the quadratic ODE and the spatial discretization scheme.
\subsection{Quadratic ODE Problem}
A $1$-$D$ quadratic ODE, when posed as an initial value problem, takes the following form. 
\begin{equation}
\frac{\d{\textbf{u}}}{\d{t}} = F_2 \textbf{u}^{\otimes 2}+F_1 \textbf{u}, \qquad
\textbf{u}(0) = \textbf{u}_{\mathrm{in}}.
\label{eq:NODE}
\end{equation}
Here $\textbf{u}=[u_1, \ldots, u_n]^{T}\in\R^n$, $\textbf{u} ^{\otimes 2}=[u_1^2, u_1u_2, \ldots, u_1u_n, u_2u_1, \ldots, u_nu_{n-1}, u_n^2]^{T}\in\R^{n^2}$, each $u_j = u_j(t)$ is a function of $t$ on the interval $[0,T]$ for $j\in\range{n}\coloneqq\{1,\ldots,n\}$, $F_2\in\R^{n\times n^2}$, $F_1\in\R^{n\times n}$ are time-independent matrices. For a matrix, assume the symbol $\|\cdot\|$ denotes the spectral norm, while for a vector it represents the $l_2$-norm. 

\subsubsection{Quantum version of the problem}
A somewhat modified version of the problem is solved with a quantum algorithm. This concise formulation of the problem is the same as the one tackled by  \cite{Liu_Kolden_Krovi_Loureiro_Trivisa_Childs_2021, costa2025further}. 

\begin{prob}\label{prob:prob1}
 Consider a one-dimensional quadratic ODE as in \eq{NODE}. Assume $F_2$ and $F_1$ are $s$-sparse\footnote{A $s$-sparse matrix has at most $s$ nonzero entries in each row and column.}, $F_1$ is diagonalizable, and that the eigenvalues $\lambda_j$ of $F_1$ satisfy $\Re{(\lambda_n)} \le \cdots \le \Re{(\lambda_1)} < 0$. We parametrize the problem in terms of 
\begin{equation}
R \coloneqq \frac{\|u_{\mathrm{in}}\|\|F_2\|}{|\Re{(\lambda_1)}|}.
\label{eq:A1}
\end{equation}

We assume we are given oracles $O_{F_2}$ and $O_{F_1}$ that provide the locations and values of the nonzero entries of $F_2$ and $F_1$, respectively, for any desired row or column. We are also given the value $\|u_{\mathrm{in}}\|$ and an oracle $O_x$ that maps $|00\ldots0\rangle \in\C^n$ to a quantum state proportional to $u_{\mathrm{in}}$. 
We aim to produce an $\varepsilon$-approximate quantum state proportional to $u(T)$ (say, $|u(T)\rangle$) for some given $T>0$ within some prescribed error tolerance $\epsilon>0$.

\end{prob}

\subsubsection{History-state variant of the problem}
In \cite{Liu_An_Fang_Wang_Low_Jordan_2023}, there is a mention of an alternative variant of \ref{prob:prob1}, where they work with the problem of computing the solution vector $u(t)$ at various time steps between $t\in [0,\ T]$. Assume the time domain has $m$ equally spaced points such that $h = T/m$. The problem is defined as follows.

\begin{prob}[History State \cite{Liu_An_Fang_Wang_Low_Jordan_2023}] 
For problem \ref{prob:prob1}, instead of producing a quantum state $|u(T)\rangle$, produce a quantum state that is in a superposition of the solution vector at different time steps.
\begin{equation}
    |u_{\mathrm{evo}}\rangle \propto u(0)\ket{0} + u(h)\ket{1} + u(2h)\ket{2}+\cdots + u(mh)\ket{m}=\sum_{k=0}^{m} u(kh)|k\rangle 
\end{equation}
\label{prb:prob2}
\end{prob}
In the appendix, we observe that the algorithms for both problems are nearly identical, differing only in the initial encoding and the final (conditional) measurement step.

\subsubsection{Interpretation of the parameter $R$} 
It is worth reiterating the physical meaning of the parameter $R$ defined in problem \ref{prob:prob1}. In fact, it plays a pivotal role in defining the tractability regime for an efficient algorithm for the problem. In the seminal paper \cite{Liu_Kolden_Krovi_Loureiro_Trivisa_Childs_2021}, 
\begin{equation}
    R = \frac{\|u_{\mathrm{in}}\| \cdot\|F_2\|}{|\Re{(\lambda_1)}|}
\end{equation}
is interpreted as qualitatively similar to the Reynolds number, which characterizes the ratio of the (non-linear) convective forces to the (linear) viscous forces within a fluid. More generally, $R$ quantifies the strength of the non-linearity relative to dissipation.


\subsection{Fischer-KPP equation to the Quadratic ODE}
It is a well-known practice in numerical methods to convert a PDE to an ODE by an appropriate spatial discretization method. In our case, we outline this conversion via the central difference method as follows.
Let $1$-$D$ Fisher-KPP equation
\begin{equation}
\frac{\partial u}{\partial t} = D \Delta u + au + bu^{2},
\label{eq:RDPDE}
\end{equation}
where, $u:=u(x,t)$. Assume the domain of the problem is $x\in [0, 1]$ and  $t\in [0, T]$.

\makebox[0pt][l]{%
\begin{minipage}{\textwidth}
\centering
    \includegraphics[width=.50\textwidth]{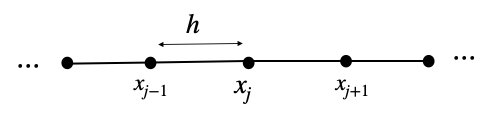}
 \captionof{figure}{Three point stencil method for special discretization }
 \label{fig:fig1}
\end{minipage}
}

On defining $u_j := u(x_j, t)$ for $j\in [n]_1$, we have $ \textbf{u} =[u_1, \ldots, \textbf{u} _{n}]^{T}\in\R^{n}$ and $ \textbf{u} ^{\otimes 2}=[u_1^2, u_1^{}u_2, \ldots, u_{n}^{}u_{n-1}, u_{n}^2]^{T}\in\R^{n^2}$. The result of the discretization is a system of non-linear ODEs as follows.

\begin{equation}
\diff{\textbf{u}}{t} = DL_h \textbf{u} +  a \textbf{u} + \textbf{b}\textbf{u}^{\otimes 2},
\label{eq:GAC}
\end{equation}

Here $L_h$ is the one-dimensional discrete Laplacian operator. For homogeneous Dirichlet boundary conditions, $L_h$ is
\begin{equation}\label{eqn:L}
         L_h := {(n+1)^2} \left(\begin{array}{ccccc}
        -2 & 1 & & &  \\
         1& -2 & 1 & & \\
          & \ddots& \ddots& \ddots& \\
           & & 1& -2 & 1\\
        & & & 1 & -2 \\
    \end{array}\right)_{n\times n}.
\end{equation}

Equation \eq{GAC} can be equivalently written as:

\begin{equation}
\diff{ \textbf{u} }{t} = (DL_h +aI) \textbf{u} + \textbf{b} \textbf{u} ^{\otimes 2},
\label{eq:sucE}
\end{equation}

Comparing it with \eq{NODE} implies 
$$F_1 = (DL_h +aI)\in \R^{n\times n} $$ and $$F_2 = \bf{b}\in \R^{n\times n^2} $$
which maps $\textbf{u}^{\otimes 2}$ to $b\textbf{u}$.\par
The above analysis suggests that a numerical scheme for the quadratic ODE can be used to predict the temporal dynamics of the Fisher-KPP equation. The rest of the document mentions (i) two existing and (ii) two new quantum algorithms for this problem.

\section{Overview of our algorithms}\label{sec: overview}

\makebox[0pt][l]{%
\begin{minipage}{\textwidth}
\centering
    \includegraphics[width=.80\textwidth]{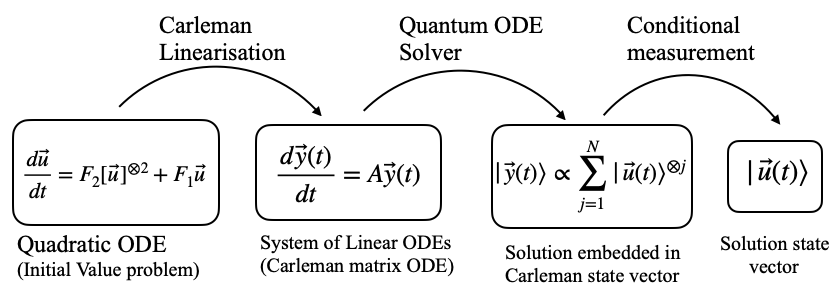}
 \captionof{figure}{Three key steps to solve the quadratic ODE problem. Existing algorithms have attempted to solve the linear ODE (step 2) in different ways. \cite{Liu_An_Fang_Wang_Low_Jordan_2023} and \cite{costa2025further} have solved it using the Euler and Taylor series methods, respectively. We have explored two different methods based on the Chebyshev series method. }
 \label{fig:fig2a}
\end{minipage}
}

The existing quantum algorithm for this problem has the following structure.
\begin{enumerate}
    \item Linearize the quadratic ODE into a system of linear ODEs\footnote{An appropriate truncation is required to deal with the emerging infinite system of linear ODEs. We will discuss it in due course. }.
    \item Develop a quantum algorithm to solve the linear ODE system.
    \item Get the desired state vector by applying a conditional measurement on the designed subsystem.
\end{enumerate}

Two linearization schemes have been primarily explored: (i) Carleman and (ii) Lie-Koopman linearization. Like \cite{Liu_An_Fang_Wang_Low_Jordan_2023} and \cite{costa2025further}, we utilize the Carleman linearization method.


\subsection{Carleman linearization for the quadratic ODE}
We elaborate on the Carleman linearization of the quadratic ODE problem. Given 
\begin{equation}
    \diff{u}{t} = F_1 u + F_2 u^{\otimes 2}, \qquad
    u(0) = u_{\mathrm{in}}.
    \label{eq:sucE}
\end{equation}

where, $u:= \vec u(t)\in \R^n$, $F_1 = (DL_h +aI)\in \R^{n\times n} $ and $F_2 = \bf{b}\in \R^{n\times n^2} $.

The Carleman embedding is given by initializing a new set of variables  $y_j: = u^{\otimes j}$ for $j = 1, 2 \cdots \infty$. It provides a system of infinite linear ODEs in the new set of variables.

\begin{equation}
\frac{\d{}}{\d{t}}
  \begin{pmatrix}
     y_1 \\
     y_2 \\
    \vdots \\
    \vdots \\
    \vdots \\
     y_{N-1} \\
     y_N \\
    \vdots
  \end{pmatrix}
=
 \begin{pmatrix}
    A_1^1 & A_2^1 &  &  &  &  &  \\
     & A_2^2 & A_3^2 & &  &  &  \\
     &  & A_3^3 & A_4^3 &  &  &  \\
     &  & \ddots & \ddots & \ddots &  &  \\
     &  &  &  & A_{N-1}^{N-1} & A_N^{N-1} &  \\
     &  &  &  &  & A_N^N & \ddots \\
     &  &  &  &  & \ddots & \ddots \\
  \end{pmatrix}
  \begin{pmatrix}
    y_1 \\
    y_2 \\
    \vdots \\
    \vdots \\
    \vdots \\
    y_{N-1} \\
    y_N \\
    \vdots
  \end{pmatrix}.
\label{eq:UODE}
\end{equation}

Where the block matrices along the diagonal position are defined as:
\begin{align}
A_1^1 &= F_1\\
A_2^2 &= (F_1\otimes I) + (I\otimes F_1) \\
A_3^3 &= (F_1\otimes I\otimes I) + (I\otimes F_1\otimes I) + (I\otimes I\otimes F_1) \\
and, A_j^j &= F_1\otimes I^{\otimes j-1}+I\otimes F_1\otimes I^{\otimes j-2}+\cdots+I^{\otimes j-1}\otimes F_1, \label{eq:tensor1}
\end{align}

Similarly, the block matrices above the diagonal are defined as:
\begin{align}
A_2^1 &= F_2\\
A_3^2 &= (F_2\otimes I) + (I\otimes F_2) \\
A_4^3 &= (F_2\otimes I\otimes I) + (I\otimes F_2\otimes I) + (I\otimes I\otimes F_2) \\
and, A_{j+1}^j &= F_2\otimes I^{\otimes j-1}+I\otimes F_2\otimes I^{\otimes j-2}+\cdots+I^{\otimes j-1}\otimes F_2, \label{eq:tensor2}
\end{align}


For developing the numerical methods, we truncate the above infinite-dimensional system of linear ODEs at a certain order $N$, thereby obtaining a finite system with the upper triangular block structure
\begin{equation}
\frac{\d{}}{\d{t}}
  \begin{pmatrix}
     y_1 \\
     y_2 \\
    \vdots \\
    \vdots \\
    \vdots \\
     y_{N-1} \\
     y_N \\
  \end{pmatrix}
=
 \begin{pmatrix}
    A_1^1 & A_2^1 &  &  &  &  &  \\
     & A_2^2 & A_3^2 & &  &  &  \\
     &  & A_3^3 & A_4^3 &  &  &  \\
     &  &  &  &  & \ddots & \ \\
     &  &  &  & \ddots &  &  \\
     &  &  &  & & & A_{N-1}^{N-1} & A_N^{N-1}   \\
     &  &  &  &  & & & A_N^N \\
     
  \end{pmatrix}
  \begin{pmatrix}
     y_1 \\
     y_2 \\
    \vdots \\
    \vdots \\
    \vdots \\
     y_{N-1} \\
     y_N \\
  \end{pmatrix}.
\label{eq:UODE}
\end{equation}

A more concise representation is as follows.
\begin{equation}
  \frac{\d{ \yy}}{\d{t}} = A  \yy, \qquad
   \yy(0) =  \yy_{\mathrm{in}}
\label{eq:LODE2}
\end{equation}

Recall, $ y_j = u^{\otimes j} $, $ \yy_{\mathrm{in}}=[u_{\mathrm{in}}, u_{\mathrm{in}}^{\otimes 2}, \ldots, u_{\mathrm{in}}^{\otimes N}]$, and $A_j^j \in \R^{n^j\times n^j}$, $A_{j+1}^j \in \R^{n^j\times n^{j+1}}$ for $j\in\range{N}$.

We will refer to the truncated Carleman matrix as \textbf{the Carleman matrix} unless stated otherwise. This finite-order truncation is bound to introduce errors. If parameter $R<1$ (eqn \eq{A1}), then it is proven that the truncation error can be arbitrarily suppressed by increasing the truncation order $N$ \cite{Liu_Kolden_Krovi_Loureiro_Trivisa_Childs_2021}.\par

Currently, there are two known quantum algorithms for the problem. A brief exposition of each algorithm is provided in the Appendix \ref{apx: apx1}. We have borrowed some of the technical machinery from them, especially the rescaling of the quadratic ODE introduced by \cite{costa2025further}. As a complementary exercise, we have conducted MATLAB simulations to solve the Fisher-KPP equation using the Taylor series method as prescribed in \cite{costa2025further}. This is to numerically validate their finding, which will ultimately provide us with a benchmark to compare our simulation results.

\section{Diagonalization of the Carleman matrix} \label{sec:diagonalization}
From the brief exposition of (i) forward Euler and (ii) Taylor series methods in Appendix \ref{apx: apx1}, it could be easily inferred that these are local approximation schemes which closely approximate the function around the initial point. Now we move to global approximation schemes, especially Chebyshev-based methods. For the ODE 
\begin{equation}
    \diff{\yy}{t} = \textbf{A}\yy(t), \qquad
    \yy(0) =  \yy_{\mathrm{in}}
\end{equation}
, there already exist two quantum ODE solvers- (i) matrix exponentiation and (ii) quantum spectral method. We will discuss their details in the next section \ref{sec: two-algo}, but the key fact about them is that both require the Carleman matrix $\textbf{A}$ to be diagonalizable. Thus, it is crucial to analyze and derive sufficient conditions for the diagonalization of the Carleman matrix.\par 

As we have seen, the Carleman matrix is not a normal matrix \footnote{See this discussion on Math StackExchange \url{https://math.stackexchange.com/a/4870629/474528}}. Thus, there is a need to exploit its intrinsic structure to infer anything conclusive about its diagonalization. In fact, we will exploit its block structure along with its sparsity to achieve this goal.\par

We start by summarizing the two central results derived in this chapter.
\begin{theorem}\label{thm:main1}
    All the eigenvalues of the Carleman matrix are real and negative if the diffusion constant $D$ and parameter $a$ are related by
    \begin{equation}
         \mathsf{4D(n+1)^2sin^2(\frac{j\pi}{2(n+1)})\geq a}
    \end{equation}
\end{theorem}
\begin{theorem}\label{thm:main2}
    The sufficient conditions for diagonalization of the Carleman matrix are (i) all the block matrices $A_j^j$ are diagonalizable, and (ii) the Carleman matrix fulfills the \textit{\textbf{No-resonance}} condition.
\end{theorem}
The next subsection deals with the proof of these two theorems.
\subsection{Sufficient condition for diagonalization of the matrix}
For the Fisher-KPP equation $\partial^t u = D\Delta_x u + au +bu^2$, we get the following quadratic ODE on $n$-point discretization, 
\begin{equation}
\diff{u}{t} = F_1 u + F_2 u^{\otimes 2}, \qquad
u(0) = u_{\mathrm{in}}.
\label{eq:sucE}
\end{equation}
where,  $F_1 = (DL_h +aI)\in \R^{n\times n} $  and  $F_2 = \bf{b}\in \R^{n\times n^2} $. Finally, Carleman linearization gives the ODE 
\begin{equation}
  \frac{\d{ y}}{\d{t}} = A  y, \qquad
   y(0) =  y_{\mathrm{in}}
\label{eq:LODE6}
\end{equation}
Now, we prove why the eigenvalues of the matrix $F_1$ are sufficient to compute the eigenvalues of the Carleman matrix $\textbf{A}$. Or, due to the upper triangular nature of the matrix $A$, the off-diagonal block matrices \textbf{don't} contribute to the eigenvalues of $\textbf{A}$.

In \cref{lem:lemma1} we derived that all the eigenvalues of the matrix $F_1$ are given by
    \begin{equation}
        \mathsf{\lambda_j = -4D(n+1)^2sin^2\left(\frac{j\pi}{2(n+1)}\right) + a;\ j\in \{1, ..., n\}}
        \label{eq:eig2}
    \end{equation}

We begin with observations on the spectral properties of $\bf{A}.$ Since, 

\begin{equation}\label{eq:carleman_structure}
    \textbf{A} =
    \begin{pmatrix}
    A_1^1 & A_2^1 &  &  &  &  &  \\
     & A_2^2 & A_3^2 & &  &  &  \\
     &  & A_3^3 & A_4^3 &  &  &  \\
     &  & \ddots & \ddots & \ddots &  &  \\
     &  &  &  & A_{N-1}^{N-1} & A_N^{N-1} &  \\
     &  &  &  &  & A_N^N & \ddots \\
     &  &  &  &  & \ddots & \ddots \\
  \end{pmatrix}
\end{equation}
Where the block matrices along the diagonal position are defined as:

\begin{align}
A_1^1 &= F_1\\
A_2^2 &= (F_1\otimes I) + (I\otimes F_1)\\
A_3^3 &= (F_1\otimes I\otimes I) + (I\otimes F_1\otimes I) + (I\otimes I\otimes F_1) \\
and, A_j^j &= F_1\otimes I^{\otimes j-1}+I\otimes F_1\otimes I^{\otimes j-2}+\cdots+I^{\otimes j-1}\otimes F_1, \label{eq:tensor1}
\end{align}
Similarly, the block matrices above the diagonal are defined as:
\begin{align}
A_2^1 &= F_2\\
A_3^2 &= (F_2\otimes I) + (I\otimes F_2) \\
A_4^3 &= (F_2\otimes I\otimes I) + (I\otimes F_2\otimes I) + (I\otimes I\otimes F_2) \\
and, A_{j+1}^j &= F_2\otimes I^{\otimes j-1}+I\otimes F_2\otimes I^{\otimes j-2}+\cdots+I^{\otimes j-1}\otimes F_2, \label{eq:carleman_A2}
\end{align}

\begin{lemma}\label{lem:lem_eigs}
    Assume that block matrix $A_1^1$ has eigenvalues $\{\lambda_1,\cdots,\lambda_d\}$, then the set of eigenvalues of block matrix  $\bf{A_j^j}$ is,
    $$ \lambda[A_j^j] =\left\{\sum_{k=1}^d m_{k}\lambda_k\ |\ m_k \in \{0, ..., j-1\} \; \mathrm{and} \; \sum_{k=1}^d  m_{k}= j \right\}. $$
\end{lemma}
For example, the set of eigenvalues of $A_2^2$ is $$\lambda {[A_2^2]} = \Big\{(\lambda_1 + \lambda_1 ), (\lambda_1 + \lambda_2 ), ..., (\lambda_1 + \lambda_d ), (\lambda_2 + \lambda_1 ), ..., (\lambda_d + \lambda_d )\Big\} .$$ 
$\mathsf{Proof:}$ As described earlier, $A_1^1 =F_1$. For the block matrix 
\begin{equation}
     A_j^j = F_1\otimes I^{\otimes j-1}+I\otimes F_1\otimes I^{\otimes j-2}+\cdots+I^{\otimes j-1}\otimes F_1
\end{equation}
Each of the matrices in the summation is Normal and pairwise commutes. Thus, the eigenvalues of $A_j^j$ are given as the sum of all possible combinations of the eigenvalues of the matrices in the summation \footnote{See this discussion on Math StackExchange \url{https://math.stackexchange.com/a/19468/474528} }.
The eigenvalues of a summand in \cref{eq:summand_kron} are 
\begin{equation}\label{eq:summand_kron}
    \lambda(I\otimes \cdots \otimes F_1\otimes\cdots \otimes I) = \lambda(F_1)
\end{equation}
due to the property of Kronecker's product. There are $j$ summand in \cref{eq:summand_kron}. It leads to the desired expression.
    \QEDB

\begin{theorem}\label{thm:no_reso_eigs}
    The set of eigenvalues of the Carleman matrix $A$ is given by
    \begin{equation}
         \lambda[A] =\left\{\sum_{k=1}^d m_{k}\lambda_k\ |\ m_k \in \{0, ..., N\} \; \mathrm{and} \; \sum_{k=1}^d  m_{k}\leq N \right\}.
    \end{equation}
\end{theorem}
$\mathsf{Proof:}$ Eigenvalues of a block upper triangular matrix are the union of the eigenvalues of the block matrices in the main diagonals \footnote{See this discussion on Math StackExchange \url{https://math.stackexchange.com/a/21461/474528}}. The eigenvalues of each block matrix $A_j^j$ are given in \cref{lem:lem_eigs}. Taking their union would give the desired set.\QEDB

\begin{definition}[\textbf{No-resonance condition}]
    Let the set of eigenvalues of the block matrix $A_j^j$ be denoted by $\{\lambda(A_j^j)\}$, then the matrix $A$ follows the No-resonance condition if and only if
    \begin{equation}
        \{\lambda(A_1^1)\} \cap \{\lambda(A_2^2)\} \cap...\cap \{\lambda(A_N^N)\} = \{\Phi\}
    \end{equation}
    \label{def1}
\end{definition}
Alternatively, any two block matrices along the diagonal of $\textbf{A}$ shall not have any common eigenvalues. There exists an equivalent (algebraic) definition to characterize the \textit{No-resonance} condition as below. 
\begin{definition} [\textbf{No-resonance condition}]
    Given $A_1^1$ has eigenvalues $\{\lambda_1,\cdots,\lambda_d\}$, Then the matrix $\bf{A}$ is said to follow the \textit{No-resonance} condition if $\forall i\in [d]$
    \begin{equation}\label{cond-res}
        \lambda_i \ne \sum_{j=1}^n m_j\lambda_j, \quad \forall m_j\in \{0, 1, ...\} \; \text{and} \ \text{ s.t. } 2\leq \sum_{j=1}^n m_j\leq N.
    \end{equation}
    \label{def1}
\end{definition}
This definition is mentioned in Equation 9 of \cite{wu2025quantum}. We derive a closed-form expression to determine if the Carleman matrix emerging from the Fisher-KPP equation satisfies the No-resonance condition.

\begin{theorem}
    Checking No-resonance for the Carleman matrix amounts to proving the following trigonometric equation has no integral solution,
    \begin{equation}\label{eq:no_reso_cond}
        \mathsf{-4D(n+1)^2sin^2\left(\frac{j\pi}{2(n+1)}\right) +a \neq \sum_{l=1}^{d}(m_i)\left( 4D(n+1)^2sin^2\left(\frac{l\pi}{2(n+1)}\right)+a\right)}
    \end{equation}
    where $ m_i\in \{0, 1, ..., n\}\ and \ 2\leq \sum m_i\leq N$.
    \label{thm:no_reso_check}
\end{theorem}
$\mathsf{Proof:}$ Take the value of $\lambda_j$ from \cref{lem:lemma1} and plug it in \cref{thm:no_reso_eigs} to get the above trigonometric relation. \QEDB
\subsection{Numerical evidence for the diagonalization}
In the previous subsection, we demonstrated that the diagonalization of the Carleman matrix reduces to verifying the No-resonance condition, which depends on four parameters. Two of them are Fisher-KPP equation parameters $D$ and $a$, while the remaining two are the truncation order $N$ and the number of grid points $n$. The latter two parameters are, in turn, influenced by the allowed approximation error $\epsilon$. To provide any theoretical guarantee, it is essential to ensure that the No-resonance condition is satisfied for all values of the associated parameters. We are currently unable to provide a theoretical guarantee. Meanwhile, we have conducted some numerical tests for it.

$\mathsf{\textbf{Numerical\ test:} }$ Although the No-resonance is equivalent to checking the feasibility of the equation \cref{eq:no_reso_cond}, it is not clear to prove or disprove it for all values of the parameters. Since we are designing a numerical solution, the following test for the no-resonance condition suggests that the condition likely holds for most parameter choices.

$\mathsf{\textbf{Method}:} $ We start with Fisher-KPP equation parameters (say, $D=0.2$ and $a=0.4$). Now, we check for various combinations of Carleman truncation order and the number of discretization points (grid) in space. The results of the No-resonance condition are tabulated in the following table.

\begin{center}
    \begin{tabular}{|c|c|c|} \hline 
        No. of grid point ($n$)& Carleman truncation order ($N$)& No-resonance condition \\ \hline 
        4 & 5 & Yes\\ \hline 
        8 & 3& Yes\\ \hline 
        8 & 4 & Yes\\ \hline 
        8 & 5 & Yes\\ \hline 
        16& 3& Yes\\ \hline 
        16& 4 & Yes\\ \hline 
        32& 3 & Yes\\ \hline
    \end{tabular}
    \label{tab:label_no_reso}
\end{center}

With our limited computing power, we could check it up to a modest value of $N=5$.  

$\mathsf{\textbf{Conclusion}:}$ Our numerical test is far from being rigorous, yet we could see some pattern in the No-resonance condition. It appears that the cases where the No-resonance condition fails must be relatively few. Even if it fails, we can fix it as follows. Assume for a particular choice of parameters $N$ and $n$, the No-resonance condition does not hold, then change these parameters to the next integer value. For example, if $n$ is not working, then take $n+1$ grid points instead of $n$. Simulating a system with $n+1$ grid points instead of $n$ won't have any impact on the usefulness of the numerical simulation.

Based on the limited empirical evidence, we forward the following hypothesis.

\begin{hypo}
    The Carleman Matrix satisfies the No-resonance condition for most of the choices of the truncation order $N$ and the number of discretization points $d$. 
    \label{hyp:hypothesis1}
\end{hypo}
One of the consequences of this would be that for most parameter choices, the emerging Carleman matrix will be diagonal. This paves the way for applying both the Chebyshev-based quantum ODE solvers we will discuss in Section \ref{sec: two-algo}.

\subsection{Diagonalization Technique } \label{subsec:diag-procedure}
Since the Carleman matrix is diagonalizable, it can be written as  
\begin{equation}
    A = V\Lambda V^{-1}
\end{equation}
where $\mathsf{\Lambda = diag(\lambda_1, ..., \lambda_d)}$. This section presents an explicit iterative method for constructing the matrix $V$ and its inverse $V^{-1}$. We use the following property of the block upper triangular matrix for this purpose.
\begin{lemma}\label{lem:diag_P}
    Let $P$ be a block upper triangular matrix
    \begin{equation}
       P= 
       \begin{bmatrix}
            P_{11} & P_{12}   \\
                    &  P_{22}
        \end{bmatrix}
    \end{equation}
    where block matrices $P_{1,1} \in R^{p \times p}$ and $P_{2,2} \in R^{q \times q}$ are known to be diagonalizable. Assume they satisfy $P_{11}\vec e_i = \lambda_i\vec e_i$ for $i\in [p]$ and $P_{22}\vec g_j = \mu_j\vec g_j$ for $j\in [q]$. Then, the set of linearly independent eigenvectors of the matrix $P$ is given by
    \begin{equation}
  \left\{  \begin{bmatrix}
    e_1 \\    0^q 
  \end{bmatrix}
  , ..., \begin{bmatrix}
    e_p \\ 
0^q 
  \end{bmatrix} 
  , ..., \begin{bmatrix}
    x_1 \\ 
g^1 
  \end{bmatrix} 
  , ..., \begin{bmatrix}
    x_q \\ 
g_q 
  \end{bmatrix} \right\}
    \end{equation}
where \textbf{vector} $x_j = (P_{11} -\mu_jI)^{-1}P_{12}g_j$.  The notation $0^q $ implies that the last $q$ entries of the column vector are zero.
\end{lemma}
Let the matrix $V$ diagonalize the matrix $P$ by a similarity transformation. i.e., $P= V\Lambda V^{-1}$. The matrix $T$ is constructed by stacking these linearly independent vectors in the columns as below.

\begin{equation}
    V=  \left[
    \begin{bmatrix}
    e_1 \\    0^q 
  \end{bmatrix}
   ... \begin{bmatrix}
    e_p \\ 
0^q 
  \end{bmatrix} 
  \begin{bmatrix}
    x_1 \\ 
g^1 
  \end{bmatrix} 
  ...\begin{bmatrix}
    x_q \\ 
g_q 
  \end{bmatrix} \right]
\end{equation}
$\mathsf{Proof:}$ Notice \footnote{The proof is inspired by the post on Math StackExchange post \url{https://math.stackexchange.com/a/21461/474528}} 

\begin{equation} 
       \begin{bmatrix}
            P_{11} & P_{12}   \\
                    &  P_{22}
        \end{bmatrix}
        \begin{bmatrix}
            e_i \\   
            0^q 
        \end{bmatrix}
        = \lambda_i
        \begin{bmatrix}
        e_i \\    
        0^q 
        \end{bmatrix}
\end{equation}
It implies that the eigenvectors of $P_{11}$ are post-padded with extra zeros ($0^q$) to give the whole matrix $P$ eigenvectors. Now we see how the eigenvectors of $P_{22}$ are pre-padded with a vector $x_j$ to get the remaining eigenvectors for $P$. Notice, 

\begin{equation} 
       \begin{bmatrix}
            P_{11} & P_{12}   \\
                    &  P_{22}
        \end{bmatrix}
        \begin{bmatrix}
            x_j \\   
            g_j 
        \end{bmatrix}
        = \mu_j
        \begin{bmatrix}
        x_j \\    
        g_j 
        \end{bmatrix}
\end{equation}
Solving this eigenvector equation gives $x_j = (P_{11} -\mu_jI)^{-1}P_{12}g_j$. Thus, pre-padding $x_j$ would give the remaining eigenvectors. It can be verified that these eigenvectors are also linearly independent. Due to the Spectral theorem, having a set of linearly independent eigenvectors is sufficient to diagonalize a matrix. One needs to pad these eigenvectors column-wise to get the similarity transformation matrix $V$. \QEDB

The vector $x_j$ is well defined if the expression $(P_{11} -\mu_jI)^{-1}$ is well defined. Or, the eigenvalues of $P_{11}$ don't coincide with any of the eigenvalues of  $P_{22}$. In fact, this is similar to the No-resonance condition. \cite{Tsiligiannis_Lyberatos_1989b} first observed this pattern, and recently \cite{wu2025quantum} has utilized this for building a quantum algorithm for quadratic ODE.

If the block matrix $P_{11}$ and $P_{22}$ can be diagonalized like 
\begin{equation}
    \mathsf{P_{11} = V_{P_{11}}\Lambda_{P_{11}} V_{P_{11}}^{-1}}\\
\end{equation}
and
\begin{equation}
    \mathsf{P_{22} = V_{P_{22}}\Lambda_{P_{22}} V_{P_{22}}^{-1},}
\end{equation}

then the following construction holds for the matrix $P$(let $P = V\Lambda V^{-1}$)
\begin{equation}
        \mathsf{V = 
        \begin{bmatrix}
            V_{P_{11}} & [X_j]   \\
            &  V_{P_{22}}
        \end{bmatrix}}
\end{equation}
where $[X_{j}] = [x_1 \cdots x_j]$.
The inverse of $V$ can be computed using the result on the block triangular matrix.
\begin{lemma}
    The inverse of a matrix 
    \begin{equation}
        \mathsf{V=\begin{bmatrix}
            V_{P_{11}} & [X_j]   \\
                    &  V_{P_{22}}
        \end{bmatrix}}
    \end{equation}
    is given by 
    \begin{equation}
        \mathsf{V^{-1}= \begin{bmatrix}
            (V_{P_{11}})^{-1} & -(V_{P_{11}})^{-1}[X_j](V_{P_{22}})^{-1}   \\
                    &  (V_{P_{22}})^{-1}
        \end{bmatrix}}
    \end{equation}
\end{lemma}
$\mathsf{Proof:} $ This can be verified by multiplying the matrix $V$ by $V^{-1}$ which yields the identity matrix\footnote{See this following discussion on Math StackExchange \href{https://math.stackexchange.com/a/2316569/474528}{https://math.stackexchange.com/a/2316569/474528}}. \QEDB

\subsection{The Diagonalization Algorithm}
Up until now, we have seen how to diagonalize the simplest possible case of the matrix $P$. This process can be iteratively applied for any larger block bi-diagonal matrices, such as the Carleman matrix.

\makebox[0pt][l]{%
\begin{minipage}{\textwidth}
\centering
    \includegraphics[width=.5\textwidth]{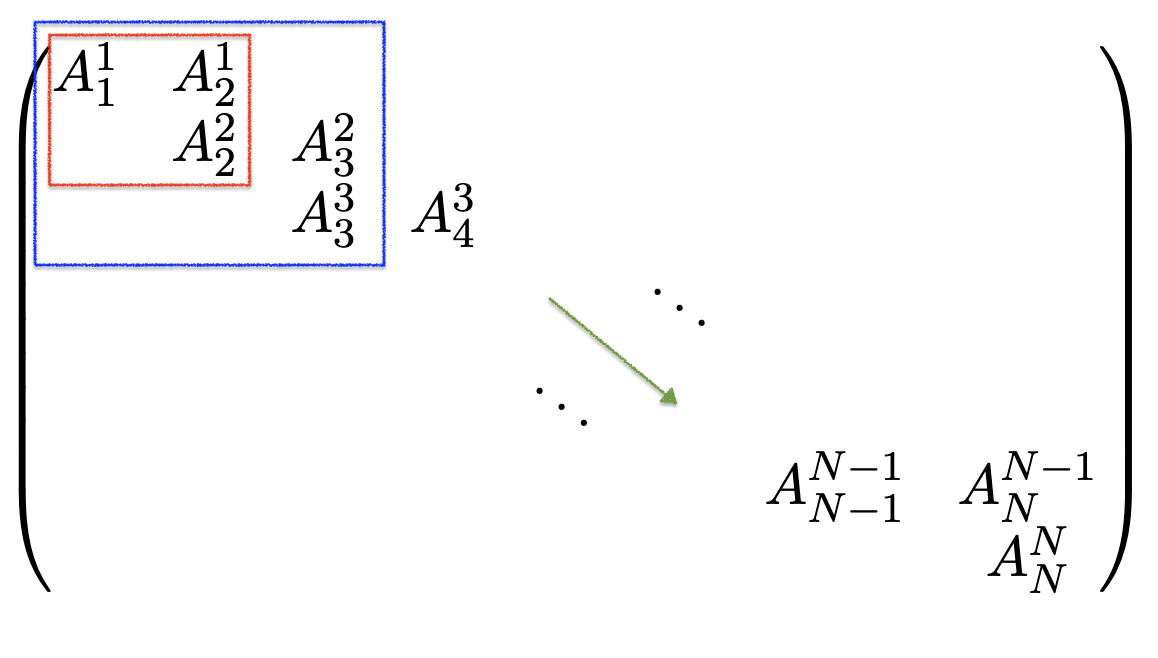}
 \captionof{figure}{(An iterative procedure for diagonalization of the Carleman matrix) It begins with the uppermost three block matrices (red) and uses \cref{lem:diag_P} to get its diagonalized form. Subsequently, it moves to the next block (blue) for diagonalization. This iterative procedure terminates at the last block $A_N^N$. }
 \label{fig:fig4}
\end{minipage}
}

The three major steps of the algorithm are as follows.
\begin{itemize}
    \item \textbf{Step I:} Start from the top three block matrices that together form a matrix like $P$ of \cref{lem:diag_P}. Since No-resonance guarantees that it can be diagonalized, we compute the associated block of the $V$ matrix.
    \item \textbf{Step II:} After having the previous block diagonalized, we add two more contiguous blocks to our analysis as shown in Fig. \ref{fig:fig4}. Again, apply \cref{lem:diag_P} to get this block diagonalized too.
    \item \textbf{Step III:} The above step is repeated till we include the last block $A_N^N$. At this point, the matrix is diagonalized.  
\end{itemize}

\makebox[0pt][l]{%
\begin{minipage}{\textwidth}
\centering
    \includegraphics[width=.7\textwidth]{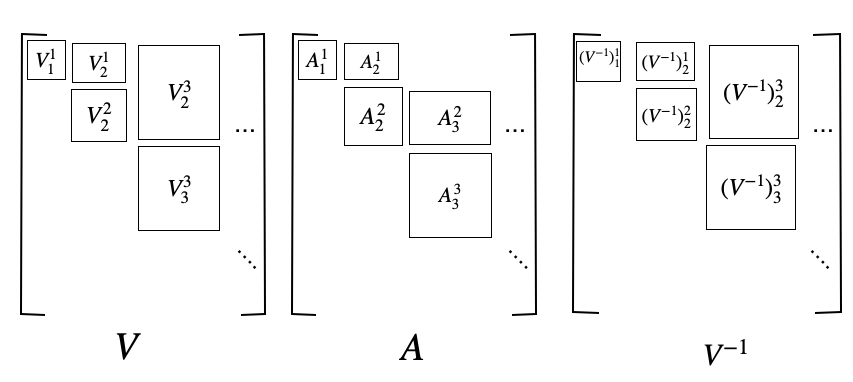}
 \captionof{figure}{(Structure of similarity transformation matrix for $A$) Although matrix $\textbf{A}$ is a block bi-diagonal matrix, the matrix $V$ and $V^{-1}$ are block upper triangular. The product of these three matrices will give the diagonal matrix $\Lambda$. }
 \label{fig:inv_carleman}
\end{minipage}
}

{\subsubsection{The computational cost of the diagonalization} }
Computing the matrix $V$ amounts to computing (i) the diagonal block $V_j^j$ and (ii) the off-diagonal block $V^j_{j+1}$. For the first part, we outline the computational procedure as follows. Given
\begin{equation}
    \mathsf{A_1^1 = F_1^1 = W\Lambda_1 W^{-1}}, \\
\end{equation}
\begin{equation}
    \mathsf{\implies \ A_2^2 = F_1\otimes I + I \otimes F_1 = (W\otimes W) \Lambda_2 (W^{-1}\otimes W^{-1})}, \\
\end{equation}
\begin{align}
    \cdots
\end{align}
\begin{equation}   
    \mathsf{\implies A_j^j = (W\otimes ...\otimes W) \Lambda_j (W^{-1}\otimes ... \otimes W^{-1}).}
\end{equation}
Thus, $V_j^j = W\otimes...\otimes W$ (Kronecker product of W with itself $j$ times), and $[V_j^j]^{-1} = W^{-1}\otimes...\otimes W^{-1}$.

For computing matrices $V^j_{j+1}$ of matrix $V$, we pre-pad the vectors $V_{j+1}^j = [X_j] = [x_1 \cdots x_j]$ as prescribed in \cref{lem:diag_P}. Computing $[X_j]$ requires computing pre-padding vector $x_{i} = (A_{i-1}^{i-1} -\mu_{i}I)^{-1} A_{i-1}^{j}g_{i}$ which in tern requires computing the inverse of $(A_{i-1}^{i-1} -\mu_{i}I)$. $(A_{i-1}^{i-1} -\mu_{i}I)^{-1}$ is equivalently written as given in the following theorem.
\begin{theorem}
    If the matrix $A_{i-1}^{i-1} = \sum_k\lambda_k g_kg_k^T$, then 
    \begin{equation}
        (A_{i-1}^{i-1} -\mu_{i}I)^{-1} = \sum_k \frac{1}{(\lambda_{k}-\mu_{i})} g_kg_k^T.
    \end{equation}
\end{theorem}
Let the computational cost of the above operation be $\mathcal{C}_{\mathrm{inv}}$. If $(A_{i-1}^{i-1} -\mu_{i}I)^{-1}$ is known, then the cost of computing $x_{i} = (A_{i-1}^{i-1} -\mu_{i}I)^{-1} A_{i-1}^{j}g_{i}$ is the same as computing two consecutive matrix-vector products. The matrix $(A_{i-1}^{i-1} -\mu_{i}I)^{-1}$ is not necessarily sparse. For the block-wise construction of matrix $V$, the most expensive step appears to be the dense upper triangular matrix-vector product.

\section{Chebyshev Series Methods for Carleman ODE} \label{sec: two-algo}
The diagonalization of the Carleman matrix, $A = V\Lambda V^{-1}$, opens a couple of new avenues for exploring an alternative solution to this problem. To the best of our knowledge, the two existing methods are- (i) Matrix exponentiation using the Chebyshev  \cite{patel_priyadarshini17}. It requires the matrix $V$ and $V^{-1}$ as part of the algorithm and (ii) Chebyshev spectral method \cite{Childs_Liu_2019} \footnote{Some literature names it as pseudo-spectral method, as they reserve the name spectral method for a slightly different procedure that also employs Chebyshev polynomials \cite{Shen_Tang}.}. It doesn't explicitly require $V$ or $V^{-1}$ in the algorithm. However, a bound on the condition number of the matrix $V$ is needed to complete the gate complexity.
We provide a detailed analysis of each method individually in the following two subsections.
\subsection{Solving Carleman ODE using Matrix exponentiation}
Since we are looking for an algorithm for the Fisher-KPP equation 
\begin{equation}
\frac{\partial u}{\partial t} = D \Delta u + au + bu^{2},
\label{eq:RDPDE2}
\end{equation}
where the domains of the parameters are $x\in [0, 1]$ and  $t\in [0, T]$, we convert it into the quadratic ODE problem upon spatial discretization. \cite{Costa_Schleich_Morales_Berry_2023} has demonstrated that scaling the Carleman matrix offers an advantage (See Appendix \ref{apx: apx1-2}). The details of the scaling procedure are as follows.
\subsubsection{Scaling the Quadratic ODE}
Scaling enhances the success probability of measuring the output state $u(T)$ when we perform the conditional measurement on the Carleman vector $\yy(T)$. 

\begin{definition}[Rescaled Quadratic ODE problem]\label{defn:rescaled-carleman}
     Consider a nonlinear ODE system of the form $\diff{\uu}{t} = F_1 u + F_2 \uu^{\otimes 2}$ as in \cref{eq:sucE}. Then, using a variable transformation in the form of a rescaling $\widetilde{u}= u/\gamma$ with $\gamma>0$, we obtain another system in the rescaled variable 
    \begin{equation}
    \label{eq:resc_ODE}
        \diff{\widetilde{\uu}}{t} = \widetilde{F}_1 \widetilde{\uu} + \widetilde{F}_2 \widetilde{\uu}^{\otimes 2} , \quad 
    \end{equation}
    with $\widetilde{F}_1 = F_1$ and $\widetilde{F}_2 = \gamma F_2$. \QEDB
\end{definition}
\textbf{Good choice of the scaling factor:} Equation 55 of \cite{Costa_Schleich_Morales_Berry_2023} prescribes taking the scaling factor as
\begin{equation}
    \mathsf{\gamma \leq \frac{||u_{\mathrm{in}}||}{R}}.
\end{equation}
It implies $   \|\widetilde{\uu}(t)\|  <1$ if $R<1$.  

\textbf{Scaled Carleman ODE:} The scaling factor slightly alters the Carleman ODE problem. We get 
\begin{align}
    \widetilde{A}_j^{j} = A_j^{j}\ and\ \widetilde{A}^{(j)}_{j+1} = \gamma A^{j}_{j+1} \\
    \widetilde{\yy} = [\widetilde{u},\ \widetilde{u}^{\otimes 2}, \cdots,\ \widetilde{u}^{\otimes N}]
\end{align}
As a result, we can write the rescaled Carleman linearization as
\begin{equation}
\dt{\widetilde{\yy}} = \widetilde{{A}}\widetilde{\mathbf{y}}, 
\end{equation}
where 
\begin{equation}\label{eq:resccarlmatrix}
\widetilde{A}=
\begin{bmatrix}
A_1^{1} & \gamma A^{1}_{2}          & \cdots & 0        & & 0             & \cdots       & 0          \\
0         & A_2^{2}  & \gamma A^{2 }_{3} & 0        & 0          & 0  & 0            & \vdots     \\
\vdots    & 0          & \ddots &          &            & 0             & \ddots       & 0          \\
          &            & \ddots & \ddots   &            &               & \ddots       & 0  \\
          &            &        &          & \ddots     &               &              & 0          \\
          &            &        &          & \ddots     & \ddots        &              & \vdots     \\
\vdots    &            &        &          &            & 0             & A_{N-1}^{N-1 }& \gamma A_{N-1}^{N}          \\
0         & 0          & \cdots &          &            & \cdots        & 0            & A_N^{N }  \\
\end{bmatrix} .
\end{equation}

\textbf{Two important results:} We derive an important result of measuring $\Tilde{u}(T)$ by conditional measurement on $\widetilde{\yy}(T)$. Recall that the (rescaled) Carleman vector would be
\begin{equation}
    \widetilde{\yy}(t) = [\widetilde{\uu}(t), \widetilde{\uu}^{\otimes 2}(t),..., \widetilde{\uu}^{\otimes N}(t)]
\end{equation}

    Thus 
    \begin{equation}
        \|\Tilde{\yy}(t)\|^2 = \sum_{l=1}^N \|\Tilde{\uu}(t)\|^l = \|\Tilde{\uu}(t)\|^2 \left(\frac{1-\|\Tilde{\uu}(t)\|^{2N}}{1-\|\Tilde{\uu}(t)\|} \right)
    \end{equation}
For  $   \|\widetilde{\uu}(t)\| <1$, 
\begin{equation}
    \left(\frac{1-\|\Tilde{\uu}(t)\|^{2N}}{1-\|\Tilde{\uu}(t)\|} \right)<N
\end{equation}
Thus, 
\begin{equation}\label{eq: bound_carl}
       \mathsf{ \|\Tilde{\yy}(t)\|^2 < \|\Tilde{\uu}(t)\|^2  N}
\end{equation}
 It is a remarkable result as it relates the norm of the Carleman vector to the norm of $\Tilde{\uu}(t)$ (the solution of the quadratic ODE ).  
The probability of measuring $\Tilde{\uu}(t):=\Tilde{\yy}_1(t)$
\begin{equation}\label{eq:prob_main}
   \mathsf{ P(\Tilde{\uu}(t)) = \frac{\|\Tilde{\uu}(t)\|^2}{\|\Tilde{\yy}(t)\|^2}\geq \frac{1}{N}}
\end{equation}
Thus, $O(1/(\sqrt{P(\Tilde{\uu}(t))}))$ rounds of amplitude amplification would boost the chance of measuring $\Tilde{\uu}(t)$ to nearly 100 percent. 
We will extensively use \cref{eq:prob_main} and \cref{eq: bound_carl} in the analysis of the runtime of the algorithms. For convenience, we will drop the tilde (symbol) above the matrix $A$ and vector $\yy$.

\subsection{Method I: Matrix Exponentiation Method}
The Carleman ODE 
\begin{equation}
  \frac{\d{\yy}}{\d{t}} = A \yy, \qquad
  \yy(0) = \yy_{\mathrm{in}}
\label{eq:LODE7}
\end{equation}
is a homogeneous first-order matrix ODE with a time-independent matrix coefficient. Thus, a formal solution exists as follows\footnote{See Chapter 3 \cite{Moya-Cessa_Soto-Eguibar_2011} for a closed-form solution of such an ODE.}.
\begin{equation}
    \mathsf{\yy(t) = e^{At}\yy(0)}
\label{eq:MExp}
\end{equation}

For specific structured Hermitian matrices, \cite{patel_priyadarshini17} introduced a quantum ODE solver that exploits the following two facts- (i) the expression $e^{\mathbf{A}t}$ can be approximated by the Chebyshev series, and (ii) a potential quantum speed-up is possible for the sparse matrix-vector multiplication. \par 

Fora  sparse Hermitian matrix $H$, a faster quantum algorithm does exist to numerically approximate the expression $e^{-Ht}\textbf{b}$. The matrix $H$ can be viewed as an operator in the Hilbert space. Section 4 of \cite{patel_priyadarshini17} mentions three requirements on $H$ for designing an efficient algorithm. First, the Hilbert space is a tensor product of many small components. Second, the components have only local interactions that make the matrix $H$ sparse. Third, both the matrix $H$ and the vector $b$ are specified in terms of a finite number of efficiently computable functions. Or, equivalently, oracles for matrix $H$ and vector $b$ are provided.\par

We will briefly see that these requirements are easily met by the Carleman matrix $A$. Thus, we apply this quantum ODE solver and analyze its overall complexity.

\textbf{Extending the algorithm for the Carleman ODE:} The Carleman matrix $\textbf{A}$ has the following properties. First, it is a non-Hermitian matrix. Second, it is $3N$-sparse matrix with local structure (see equation \cref{eq:carleman_structure} to equation \cref{eq:carleman_A2}). Third, there is an explicit construction of the matrix $V$ (i.e., $A= V\Lambda V^{-1}$) guaranteed by the No-resonance condition. \par

We deploy the algorithms by \cite{patel_priyadarshini17} for the Carleman matrix. To this end, we state the three crucial lemmas.
\begin{lemma}\label{lem:scale_cheby}
    The Carleman matrix $A$ can be linearly transformed to map its eigenvalues to the interval $[-1,\ 1]$.
    \begin{equation}
        A \mapsto  \frac{A -  \frac{(\lambda_{\mathrm{max}} + \lambda_{\mathrm{min}})}{2}I}{\frac{\lambda_{\mathrm{max}} - \lambda_{\mathrm{min}}}{2}}
    \end{equation}
\end{lemma}

Let's define $\alpha := \frac{(\lambda_{\mathrm{max}} + \lambda_{\mathrm{min}})}{2}$ and $\beta:= \frac{(\lambda_{\mathrm{max}} - \lambda_{\mathrm{min}})}{2}$. The scaling of time goes as 
\begin{equation}
    t\mapsto \beta t
\end{equation}

It should be noted that the above scaling doesn't alter the diagonalization procedure due to the following lemma. 
\begin{lemma}\label{lem:diag_cheby}
    The Carleman matrix $A$ is diagonalized as $A=V\Lambda V^{-1}$, then the same similarity transformation matrices $V$ and $V^{-1}$ will diagonalize the scaled matrix. 
    \begin{equation}
        A \mapsto  \frac{A -  \frac{(\lambda_{\mathrm{max}} + \lambda_{\mathrm{min}})}{2}I}{\frac{\lambda_{\mathrm{max}} - \lambda_{\mathrm{min}}}{2}}
    \end{equation}
\end{lemma}
Proof: Notice
\begin{equation}
    \mathsf{V(A/\beta - \alpha I/\beta) V^{-1}= V(A/\beta)V^{-1}  - V(\alpha I/\beta) V^{-1} = \Lambda /\beta - \alpha I/\beta}.
\end{equation}
This is also a diagonal matrix. \QEDB

Consequently, we use the same matrices $V$ and $V^{-1}$ without worrying about the scaling.

\begin{lemma}\label{lem:eigs_carl}
    The minimum and maximum values of the Carleman matrix are
    \begin{equation}
        \mathsf{\lambda_{\mathrm{max}} = -4D(n+1)^2sin^2\left(\frac{\pi}{2(n+1)}\right) + a}
        \label{eq:min_eigs}
    \end{equation}
    \begin{equation}
        \mathsf{\lambda_{\mathrm{min}} = -N\left(4D(n+1)^2sin^2\left(\frac{n\pi}{2(n+1)}\right) + a\right)}    
    \label{eq:min_eigs}
    \end{equation}
\end{lemma}

$\mathsf{Proof}$: We work in a parameter regime with negative eigenvalues. Thus, the maximum eigenvalue of the matrix $A$ is the one with the {minimum magnitude} eigenvalue of $A_1^1 = F_1$. The maximum eigenvalue of the matrix $A$ is the one with the {maximum magnitude} eigenvalue of $A_N^N$. \QEDB

Since $A =V\Lambda V^{-1}$, we express the analytic equation of the Carleman matrix as 
\begin{equation}
    \mathsf{e^{At} = Ve^{\Lambda} V^{-1} }.
\end{equation}
Here, 
\begin{equation}
e^{\Lambda t}=
    \begin{pmatrix}
    e^{\lambda_1 t} &  &  &  &  \\
     & e^{\lambda_2t} &  &  &  \\
     & & \ddots &  &  \\
     &  &  & e^{\lambda_{n-1} t} &  \\
     &  &  &  & e^{\lambda_{d} t} \\
\end{pmatrix}.
\end{equation}

Express each of the exponential functions along the diagonal as a Chebyshev series
 \begin{equation}
         \mathsf{e^{\lambda_i t} = \sum^{\infty}_{k=0}C_k(t)T_k(\lambda_i)}
\label{eq:cheb_series}
\end{equation}
These are two crucial results regarding the series that is used to design and analyze the algorithm.
\begin{theorem}\label{lem:bessel}
    In the Chebyshev series 
    \begin{equation}\label{eq:cheb_series}
        \mathsf{e^{\lambda_i t} = \sum^{\infty}_{k=0}C_k(t)T_k(\lambda_i)}
    \end{equation}
    The coefficients are given by the modified Bessel function of the first kind $I_k(x)$:
    \begin{equation}
        \mathsf{C_0 = \frac{1}{\pi} \int_0^{\pi}e^{-tcos\theta}d\theta = I_0(t),}
    \end{equation}
    \begin{equation}
        \mathsf{C_{k>0} = \frac{2}{\pi} \int_0^{\pi}e^{-tcos\theta} cos(k\theta)d\theta = 2(-1)^kI_k(t)}    
    \end{equation}
Where, 
\begin{equation}
    \mathsf{I_{k(t)} = \sum_{m=0}^{\infty} \frac{(t/2)^{k+2s}}{s!(k+s)!}}
\end{equation}
\end{theorem}
See Mathews \& Walker's book \cite{mathew1971} for its derivation using complex analysis. 
\begin{theorem}[\cite{patel_priyadarshini17}] 
    Given a simulation time of $t$ and an error tolerance of $\varepsilon$, the maximum truncation order is given by 
    \begin{equation}
       \mathsf{ r = \frac{e^{5/4 t}}{2} + ln(1/\varepsilon) = O(t+log(1/\varepsilon))}
    \end{equation}
    \label{thm:truncate_cheby}
\end{theorem} 
The proof is proved in the paper \cite{patel_priyadarshini17} and it follows as a natural consequence of \cref{lem:bessel}.\par
Now we outline the steps of the algorithm as follows.

\textbf{Step I:} Rescale the Carleman matrix such that its eigenvalues lie in $[-1,\ 1]$, which is the domain of Chebyshev polynomials. The method for rescaling is given in \cref{lem:scale_cheby}. 

\textbf{Step II:} Construct the matrix $V$ and $V^{-1}$ using the method described in subsection \ref{subsec:diag-procedure}.

\textbf{Step III:} Based on user input of allowed error $\varepsilon$ and simulation time, truncate the series up to order $r$ using \cref{thm:truncate_cheby}. It gives
\begin{equation}\label{eq:cheby_approx}
    \mathsf{e^{\lambda_i t} \approx [e^{\lambda_i t}]_r = \sum^{r}_{k=0}C_k(t)T_k(\lambda_i)}.
\end{equation}

For convenience, we use the following notation for truncated series
\begin{equation}
    \mathsf{exp(\Lambda T) \approx [exp(\Lambda T)]_r}.
\end{equation}

\textbf{Step IV:} Perform the quantum implementation of 
\begin{equation}
    \mathsf{\yy(T) = V[exp(\Lambda t)]_r V^{-1}\yy(0)}
\end{equation}
for determining the solution at some given time $T$.

\textbf{Step V:} After $O(1/(\sqrt{P(\Tilde{\uu}(t))})) = O(\sqrt{N})$ rounds of amplitude amplification (see \cref{eq:prob_main}), the conditional measurement on $\yy(T)$ would give $\uu(T)$ with high probability.

\makebox[0pt][l]{%
\begin{minipage}{\textwidth}
\centering
    \includegraphics[width=.5\textwidth]{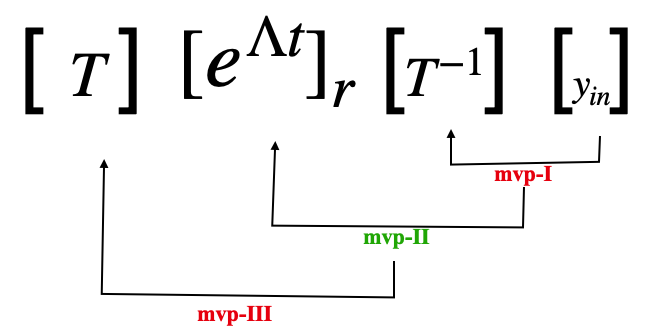}
    
 \captionof{figure}{Step IV involves three matrix-vector products (mvp) where matrices can be viewed as operators acting on a state vector. A known quantum advantage exists in the sparse matrix-vector structure. The mvp-II is efficient (see \cref{thm:pp_main}). However, it is unclear whether we can achieve similar efficiency in computing mvp-I and III, where matrices $V$ and $V^{-1}$ are (dense) upper triangular matrices.}
 \label{fig:fig2a}
\end{minipage}
}

\QEDB

There are two crucial results for estimating the cost of the matrix-vector product employed in the algorithm.
\begin{theorem}[\cite{patel_priyadarshini17}]
Let $d$ be the size of the Carleman matrix. The error tolerance is $\varepsilon$, and the Chebyshev truncation order is $r$. The quantum cost of implementing the matrix-vector product (II), i.e., $[e^{\Lambda t}]\textbf{b}$, is
    \begin{equation}
        \mathsf{O\left(log(d)\left(t+ log\left(\frac{ 1 }{\varepsilon}\right)\right) log^{3}\left(         \frac{(t+log(1/\varepsilon))}{e^{-t} \varepsilon}\right)\right)}
    \end{equation}
    \label{thm:pp_main}
\end{theorem}
$\mathsf{Proof:}$ We get this using equation 59 of \cite{patel_priyadarshini17}, which provide an estimate for the sparse matrix-vector multiplication. That paper uses the symbol $d$ to represent sparsity and $ N = 2^n$ to denote the matrix size. In our case, the sparsity of the diagonal matrix $[e^{\Lambda t}]_r$ is one while the size of the Carleman matrix is $d$. Thus substituting $\mathsf{n=log(d)}$ in equation 59 of \cite{patel_priyadarshini17} gives the desired result.\QEDB

\begin{theorem}
    The computational cost of multiplying an upper triangular matrix $M$ of size $d\times d$ with a vector $\textbf{b}$ is
    \begin{enumerate}
        \item Quantum cost: $\mathsf{O(d\cdot log(d))}$
        \item Classical cost: $\mathsf{O(d^2)}$
    \end{enumerate}
    \label{thm:costs}
\end{theorem}
$\mathsf{Proof:}$ In the classical case, the vector $\textbf{b}$ is multiplied by each $d$ row of $M$. Each such operation requires $d$ multiplication and addition. Hence, the complexity of $\mathsf{O(d^2)}$. \par In the quantum case, for each $d$ row and vector $b$, multiplication requires $\mathsf{log(d)}$ instructions. It is similar to the single-instruction-multiple-data (SIMD) paradigm of parallel computing. Here, the same instruction (gates) is given to the qubits (data) in superposition. See \cite{patel_priyadarshini17} for an elaborate discussion. \QEDB

\subsubsection{Computational cost of the algorithm}
Each of the steps has the following complexity.
\begin{itemize}
    \item Step I (matrix rescaling): The matrix rescaling has a relatively nominal cost because there is a closed-form expression for computing the minimum and maximum eigenvalue of the Carleman matrix due to \cref{lem:eigs_carl}.
    
    \item Step II (diagonalization): The cost of finding the matrix $V$ and $V^{-1}$ is discussed in Chapter \ref{sec:diagonalization}. The most expensive step is to compute the off-diagonal block elements. (See \cref{fig:inv_carleman}). We don't immediately see any advantage in computing this step on a quantum computer. Thus, it can be computed on a classical computer as part of classical pre-processing.
    
    \item Step III (finding the truncation order): Estimating the Chebyshev series truncation order $r$ has a relatively nominal cost.
    
    \item Step IV (matrix-vector product): As per \cref{thm:costs} and \cref{thm:pp_main}, the leading order contribution of various parameters in the gate complexity is $\mathsf{O(t\cdot polylog(t))}$, $\mathsf{O(d\cdot log(d))}$, and $\mathsf{O(polylog(1/\varepsilon))}$. \textbf{Note:} This will output the quantum Carleman vector $\ket{\yy(T)}$.
    \item Step V (amplitude amplification): As per \cref{eq:prob_main},  $O(1/(\sqrt{P(\Tilde{\uu}(t))}))=O(\sqrt{N})$ rounds of amplitude amplification would boost the change of measuring $\Tilde{\uu}(t)$ (by the conditional measurement on $\yy(T)$) to nearly 100 percent.
\end{itemize}

To conclude, we present Method I as a hybrid Classical-Quantum algorithm. The scaling of gate complexity with respect to parameters $T$ and $\varepsilon$ is satisfactory (See discussion \ref{sec: summary}). The scaling of the gate complexity with matrix size is $\mathsf{d(log(d))}$. Currently, we can't find any local sub-structures in the matrix $V$ that a quantum algorithm could harness to reduce the gate complexity poly-logarithmic in $d$.

This concludes the theoretical analysis of the first algorithm. Next, we proceed to the second algorithm for the problem.
\subsection{Method II: Quantum Spectral Method}
The high-level description of the algorithm is very similar to that of other Quantum linear system-based algorithms. The key difference is that the Carleman ODE is solved using the Chebyshev polynomial-based method.

\label{sec:main}
\begin{itemize}
    \item  We state the main theorem of \cite{Childs_Liu_2019}. This is presented within the broader context of a first-order matrix differential equation. If certain conditions are fulfilled, the theorem provides quantum gate complexity for solving such an ODE.
    \item We then prove why the Carleman ODE satisfies all the theorem requirements. Thus, we can use the Pseudo-spectral algorithm that outputs the desired quantum encoding of the solution $\ket{\yy(T)}$.
    \item We use the conditional measurement to get $\ket{\uu(T)}$ from the Carleman vector $\ket{\yy(T)}$.
\end{itemize}

\begin{theorem}[Adapted from theorem 1 of \cite{Childs_Liu_2019}]
Consider the following homogeneous ODE problem.
\begin{equation}
    \diff{\yy}{t} = A\yy;\ \ \yy(0) = \yy_{\mathrm{in}} 
\end{equation}

Assume a $s$-sparse matrix $A\in \R^{d\times d} $ can be diagonalized as $A=V\Lambda V^{-1}$ where $\Lambda=diag(\lambda_1,\ldots,\lambda_d)$ with $\Re(\lambda_i)\le0$ for each $i \in d$. Let the condition number of the matrix $V$ be $\kappa_V$. Then, there exists a quantum algorithm that produces $\epsilon$-close quantum state $|\yy(T)\rangle/\| \yy(T)\|$  succeeding with probability $\Omega(1)$ using
\begin{equation}
    \mathsf{O\left(\kappa_Vs\|A\|Tq \poly\left(\log\left(\frac{\kappa_V s \|\yy_{\mathrm{in}}\|\|A\|T}{\epsilon g}\right)\right)\right)}
\end{equation}
queries to oracles $O_{A}$ and $O_{\yy_{\mathrm{in}}}$. And parameters 
\begin{align}
g&:=\|\yy(T)\|, &
q&:=\max_{t\in [0,T]} \frac{\|\yy(t)\|}{\|\yy(T)\|}.
\label{eq:main_thm_parameters}
\end{align}
The overall gate complexity is larger than the query complexity by a factor of
\begin{equation}
    \poly \left(\log \left(\frac{\kappa_Vs(d)\|A\|\|\yy_{\mathrm{in}}\|T}{\epsilon}\right) \right)
\end{equation}
\label{thm:main}
\end{theorem}

$\mathsf{Proof}$: This follows from Theorem 1 of \cite{Childs_Liu_2019} , which is written in the context of a time-dependent matrix $A(t)$. In our case, the matrix $A$ is time-independent. Thus, we estimate an upper bound on the parameter 
\begin{equation}
    g':=\max_{t\in [0,T]} \max_{n \in \N}\|\hat{x}^{(n+1)}(t)\|
\end{equation}
because
\begin{equation}
    \max_{t\in [0,T]} \max_{n \in \N}\|\hat{x}^{(n+1)}(t)\| \leq \kappa_V \|\yy_{\mathrm{in}}\|.
\end{equation}
The last result is inferred from Lemma 3 and equation 4.21 in \cite{Childs_Liu_2019}. \QEDB

As we have seen, the Carleman Matrix fulfils the requirements mentioned in the theorem.
\begin{itemize}
    \item All its eigenvalues are non-positive (due to \cref{thm:main1})
    \item It is Diagonalizable ($A=V\Lambda V^{-1}$) (due to \cref{thm:main2})
    \item It is ($3N$)-sparse, where $N$ is the truncation order and $s$ is the sparsity of the matrix $F_1$. (due to \cref{lem:sparsity_carleman})
\end{itemize}
The gate complexity requires computing the following problem parameters in terms of the parameters of the quadratic ODE problem.
\begin{itemize}
    \item Spectral norm : $\|A\|$
    \item $s$: sparsity of $A$
    \item $d$ : the size of $A$
    \item condition number $\kappa_V$
    \item $g:=\|\yy(T)\|$, and
    \item $q:=\max_{t\in [0,T]} \frac{\|\yy(t)\|}{\|\yy(T)\|}.$
\end{itemize}
We estimate them as follows:

\textbf{The estimation of the spectral norm of $A$:} $\|A\|$ can be simplified in terms of system parameters. In the context of the reaction-diffusion equation, \cite{Liu_An_Fang_Wang_Low_Jordan_2023} has estimated $\|A\|$ as follows.
\begin{lemma}[Equation 4.19 and 4.20 in $\cite{Liu_An_Fang_Wang_Low_Jordan_2023}$]
    For the Carleman matrix $A$, its spectral norm is bounded by 
    \begin{equation}
        \|A\|\leq N (\|F_1\|+\|F_2\|)
    \end{equation}
    The spectral norm of $F_1$ and $F_2$ are given as
    \begin{align}
        \|F_1\| = 4D(n+1)^2 +a \\
        \|F_2\| = b
    \end{align}
\end{lemma}
We estimate each of the above parameters individually as follows.

\textbf{Estimation of the sparsity of $A$:} The sparsity of the Carleman $A$ is given in terms of the sparsity of the matrix $F_1$ and the Carleman truncation order as 
\begin{lemma}[Equation 4.19 and 4.20 in $\cite{Liu_An_Fang_Wang_Low_Jordan_2023}$]
    For an $N$-th order truncated Carleman matrix $A$, the matrix sparsity is $s=3N$. 
\end{lemma}
The formal argument is provided in $\mathsf{Proof}$:  \cref{lem:sparsity_carleman}. The sparsity of the matrix $F_1 = 3L_h+aI$ is 3 because of the sparsity of the discrete Laplacian matrix.

\textbf{The size of the matrix $A$:} It is given by the lemma \cref{lem:size}, and scales scales as $d=O(n^N)$, where $n$ is the number of discretization points, while $N$ is the Carleman truncation order.

\textbf{Estimating parameter $g$:} We estimate it using the \cref{eq: bound_carl}. Since, $g:=\|\yy(T)\|$, putting $t=T$ in \cref{eq: bound_carl} implies
\begin{equation}
    \mathsf{g:=\|\yy(T)\|\leq \|\uu(T)\|\sqrt{N}}
\end{equation}

\textbf{Estimating parameter $q$:} We again estimate it using the \cref{eq: bound_carl}. Since,

\begin{equation}
    q:=\max_{t\in [0,T]} \frac{\|\yy(t)\|}{\|\yy(T)\|}\leq \max_{t\in [0,T]} \frac{\|\uu(t)\|}{\|\uu(T)\|}
\end{equation}
Since the system is dissipative (negative eigenvalues), $\uu(t)<\uu_{\mathrm{in}}$. Combining it with the above results gives 
\begin{equation}
    \mathsf{q:=\max_{t\in [0,T]} \frac{\|\yy(t)\|}{\|\yy(T)\|}\leq \max_{t\in [0,T]} \frac{\|\uu(t)\|}{\|\uu(T)\|}\leq \frac{\|\uu_{\mathrm{in}}\|}{\|\uu(T)\|}}
\end{equation}

\textbf{Compiling the results:} We put all the above-estimated quantities into the gate complexity of \cref{thm:main}. It implies the number of Oracle calls to $O_A$ and $\yy_{\mathrm{in}}$

\begin{equation}
    O\left(\kappa_V(3N)[N(4D(n+1)^2+a +b)] T \frac{\|\uu_{\mathrm{in}}\|}{\|\uu(T)\|} \poly\log\left(\frac{\kappa_V (3N) (\|\uu_{\mathbf{in}}\| \sqrt{N}) [N(4D(n+1)^2+a+b)] T}{\epsilon \uu(T)}\right)\right).
\end{equation}
While the gate complexity is larger by a factor of 

\begin{equation}
    \mathsf{\poly \log \left(\frac{\kappa_V(3N)(n^N)[N(4D(n+1)^2+a+b] \|\uu_{\mathrm{in}}\|\sqrt{N} T}{\epsilon}\right)}
\end{equation}
\textbf{Overall scaling: } The leading order contribution of various parameters in the gate complexity is $\mathsf{O(T\cdot polylog(T))}$, $\mathsf{O(N^2polylog(N^{1.5}n^N))}$, and $\mathsf{O(polylog(1/\varepsilon))}$.

\textbf{Remark:} A bound on the condition number of $V$ in terms of system parameters would also be very useful, as it appears in the expression on the gate complexity. We have not found a conclusive upper bound, but we have outlined our attempt in Appendix \ref{apx: apx2-3}.

For the sake of completeness, we briefly summarize the Pseudo-spectral method. The key goal is to solve the homogeneous ODE 
\begin{equation}
  \frac{\d{\yy}}{\d{t}} = A \yy, \qquad
  \yy(0) = \yy_{\mathrm{in}}
\label{eq:LODE5}
\end{equation}
The goal is to find $\yy(t)$ for $t\in[0,\ T]$. 

\textbf{Key idea: }
Consider a truncated Chebyshev series approximation for the solution $\yy(t)$ as
\begin{equation}
    \yy_i(t)\approx \hat{\yy}_i(t)=\sum_{k=0}^r c_{i,k}T_k(t),\quad i\in\rangez{d}:=\{0,1,\ldots,d-1\}
\label{eq:cheby_expand}
\end{equation}
There are two useful properties of the above consideration.
\begin{itemize}
    \item Special points exist in the domain $t\in[0,\ T]$ where evaluating the Chebyshev polynomial has a marginal cost. These points are called Chebyshev-Gauss-Lobatto quadrature (CGL) nodes, $t_l=\cos\frac{l\pi}{r}$ for $l \in \rangez{r+1}$. Note the value of the Chebyshev polynomial at CGL nodes $T_k(t_l)= \cos\frac{kl\pi}{r}$.
    \item For $\forall i\in [d]_0$, the solution components $\yy_i(t)$ satisfy the ODE and initial conditions at these CGL nodes $\{t_l\}_{l=0}^n$. It would give a system of linear equations solved for the coefficients.
\end{itemize}

\textbf{Determining the Chebyshev truncation order:} The criteria for selecting the Chebyshev truncation order $r$ are due to the following theorem.

\begin{lemma}[Lemma 2 of \cite{Childs_Liu_2019}]\label{lem:smooth_spectral_approx}
Let $\yy(t)\in C^{\infty}(-1,1)$ be the analytic solution of an ordinary differential equation and let the numerical solution be $\hat{\yy}(t)$ that satisfies the ODE and initial condition for $\{t_l=\cos\frac{l\pi}{n}\}_{l=0}^n$. Then
\begin{equation}
\max_{t\in[-1,1]}\|\hat{\yy}(t)-\yy(t)\|\le \sqrt{\frac{2}{\pi}} \max_{t\in[-1,1]}\|{\yy}^{(r+1)}(t)\|\left(\frac{e}{2r}\right)^r.
\end{equation}
\end{lemma}
If the matrix $A$ is diagonalizable, $A=V\Lambda V^{-1}$, then equation 4.21 of \cite{Childs_Liu_2019} gives
\begin{equation}
\|\hat{\yy}(T)-\yy(T)\|\le m\kappa_V \|\yy_{\mathrm{in}}\| \left(\frac{e}{2r}\right)^r
\end{equation}
where, $m\geq \frac{\|A\|T}{2}$. Thus, expressing the formula below gives an estimate of the condition number in terms of time $T$ and error $\varepsilon$. 
\begin{equation}
\|\hat{\yy}(T)-\yy(T)\|\le \left(\frac{\|A\|T}{2}\right)\kappa_V \|\yy_{\mathrm{in}}\| \left(\frac{e}{2r}\right)^r
\end{equation}

To conclude, the gate complexity with respect to parameter $N$, the matrix size $d$, time $T$, and error $\epsilon$ is $\mathsf{O(\kappa_V N^2 T polylog(dTN/\varepsilon))}$. It is worth noting that a definite upper bound on $\kappa_V$ will make the actual scaling more explicit.\par

In Appendix \ref{apx: simulation}, we have done a MATLAB simulation to test the numerical stability for the approximation techniques employed at various parts of the algorithm.

\section{Summary and Discussion}\label{sec: summary}
We compare the gate complexity of all known quantum algorithms, for the Fisher-KPP equation, in terms of simulation time $T$, approximation error $\epsilon$, and Carleman truncation order $N$.

\begin{center}
    \begin{tabular}{|c|l|c|c|c|} \hline 
        S. No. & Algorithms & Time ($T$) & Error ($\varepsilon$) & Carleman order ($N$)\\ \hline 
        1. & Euler Method \cite{Liu_An_Fang_Wang_Low_Jordan_2023}& $\mathsf{T^2\cdot polylog(T)}$ & $\mathsf{({1}/{\varepsilon})\cdot polylog({1}/{\varepsilon})}$ & $\mathsf{N^3 \norm{u_{in}}^N polylog(N)}$\\ \hline
        2. & Taylor method \cite{costa2025further}& $\mathsf{T\cdot polylog(T)}$ & $\mathsf{ polylog({1}/{\varepsilon})}$ & $\mathsf{N^2 polylog(N)}$ \\ \hline
        3. & \textit{Matrix Exponentiation} & $\mathsf{T\cdot polylog(T)}$ & $\mathsf{ polylog({1}/{\varepsilon})}$& $\mathsf{Nn^N \cdot polylog(N)}$ [$\dagger$]\\ \hline
        4. & \textit{Quantum Spectral Method} & $\mathsf{T\cdot polylog(T)}$ & $\mathsf{ polylog({1}/{\varepsilon})}$ & $ \mathsf{N^2\kappa \cdot polylog(Nn^N)}$ [$\ddagger$]\\ \hline
    \end{tabular}
    \label{tab: complexity algorithms}
\end{center}

$\dagger$ \textit{Matrix Exponentiation} method is presented as a classical-quantum algorithm. For a given $d$-sized Carleman matrix, the complexity of the most expensive step is $\mathsf{O(dlog(d))}$ $\mathsf{=O(Nn^N log(n))}$. \footnote{Recall, the size of the Carleman matrix scales as $d= O(n^N)$ where $n$ is the number of discretization points and $N$ is the order of truncation.} \par
$\ddagger$ For the \textit{quantum spectral} method, the upper bound on the condition number $\kappa$ of the similarity transformation matrix $V$ is unclear \footnote{Recall, the Carleman matrix $A$ can be expressed as $V^{-1}AV=\Lambda$, where $\Lambda$ is a diagonal matrix.}.\par

From the above table, we can infer that the last three algorithms have similar gate complexity with respect to (i) simulation time $T$ and (ii) approximation error $\varepsilon$. Due to the lack of an upper bound on the condition number $\kappa$ for the fourth algorithm, a conclusive comparison in terms of the Carleman truncation $N$ is difficult. The Taylor series method has a provable (nearly) quadratic dependency on $N$. The same can't be assured for the fourth algorithm, as $\kappa$ might significantly impact the overall dependency on $N$. \par

It is natural to ask how much we can improve the complexity of the quantum algorithm for this problem. It is known that the time complexity can't be made sublinear in $T$. This barrier is due to the No-fast-forwarding theorem for the Hamiltonian simulation problem \cite{berry2017quantum}, which asserts that the worst-case time complexity is $\Omega(T)$. Thus, the algorithms are pretty close to the lower bound, and there remains a scope for (poly) logarithmic improvement.\par


In the same context of Hamiltonian simulation, the query complexity has a lower bound in precision $\varepsilon$ as \cite{berry2014exponential}
\begin{equation}
    \mathsf{\Omega\left(\frac{log (1/\varepsilon)}{log(log(1/\varepsilon))} \right) \leq O(log^c(1/\varepsilon))}.
\end{equation}
for some integer $c\geq 1$. Thus, the maximum one can hope for is a polynomial improvement. \par 
The lower bound on the condition number $\kappa$ comes from a complexity theoretic conjecture. In the seminal work on the Quantum linear system problem, \cite{harrow2009quantum} proved that the complexity should scale as $\Omega(\kappa)$ if the conjecture $\mathsf{BQP \neq PSPACE}$ holds.

\textbf{Classical hardness:} We discuss the classical hardness of the problem. It is unlikely that there can be a classical algorithm having polynomial dependency on all the parameters $T,\ \varepsilon$, and $N$. The reason is that the quadratic ODE problem is $\mathsf{BQP}$-hard. This follows from the following argument. \par
In the quadratic ODE problem
\begin{equation}
\frac{\d{\textbf{u}}}{\d{t}} = F_2 \textbf{u} ^{\otimes 2}+F_1 \textbf{u}, \qquad
\textbf{u}(0) = \textbf{u}_{\mathrm{in}}.
\label{eq:NODE8}
\end{equation}
consider the special case where $F_2 = \textbf{0}$ and $F_1$ is anti-Hermitian matrix; this would encompass the Schrödinger's equation. Thus, producing the quantum state $\ket{u(T)}$, given $\ket{u(0)}$, is as hard as the quantum Hamiltonian simulation. Thus, any efficient classical algorithm for this problem would imply $\mathsf{BQP =P}$. \QEDB

\section{Open Problems} \label{sec: open-probs}
We see four major open problems related to this work. First, we haven't provided an upper bound on the condition number of the matrix $V$, the matrix which diagonalizes the Carleman matrix via similarity transformation. This is crucial for obtaining a complete estimate of gate complexity, which is exclusively defined in terms of input parameters $n$, $T$, and $\epsilon$. \par Second, to settle the hypothesis \ref{hyp:hypothesis1} on No-resonance for the Carleman matrix. Currently, we have empirically tested it up to a certain value of $n$. We don't know how it would unfold for a larger value of $n$. \footnote{Here is a discussion on Stack Exchange \href{https://math.stackexchange.com/q/4924878/474528}{https://math.stackexchange.com/q/4924878/474528}} \par Third, to extend the diagonalization analysis beyond the quadratic ODE. In Appendix \ref{apx: apx3}, we have outline the analysis for extending it to 
\begin{equation}
    \frac{\d{\textbf{u}}}{\d{t}} = F_M\textbf{u}^{\otimes M}+F_1\textbf{u}, \qquad
    \textbf{u}(0) = \textbf{u}_{\mathrm{in}}.
\end{equation}
But it is not apparent how we can approach the mixed case of 
\begin{equation}
    \frac{\d{\textbf{u}}}{\d{t}} = \sum_{k=2}^MF_k\textbf{u}^{\otimes k}+F_1\textbf{u}, \qquad
    \textbf{u}(0) = \textbf{u}_{\mathrm{in}}.
\end{equation}
This does not seem amicable to our iterative diagonalization procedure. A suitable modification needs to be made to the procedure to harness the newly emerging structure of the Carleman matrix in the new case.
\section*{Data Availability}
The MATLAB codes for the classical simulation of the quantum algorithms are available at the GitHub repository [\href{https://github.com/108mk/Quantum_Tech_Project_M_Tech_IISc/tree/f4a6f80e24864f84c451d418116c44edb6a8972d/codes}{link}].

\section*{Acknowledgments}
I thank Apoorva D. Patel for (i) his suggestion to try out the Chebyshev polynomial-based numerical method for solving the problem, (ii) for suggesting the reference on Carleman linearization with no dissipation condition by Wu et. al. (2024), and (iii) discussions on various parts of the work. \par I acknowledge that the code written by Herman Kolden for \href{https://github.com/hermankolden/CarlemanBurgers}{viscous Burgers equation} has been repurposed to reproduce the results on the forward Euler method (Appendix \ref{apx: apx1-1}). The simulation was performed on computing facilities provided by \href{https://iqti.iisc.ac.in/}{ ({IQTI, IISc Bangalore})}.

\section{References}
\bibliographystyle{alpha}
\bibliography{references}

\section{Appendix I: Survey of two existing algorithms}\label{apx: apx1}

We briefly summarize the existing methods to solve the Carleman ODE. They will serve as a benchmark for our algorithms.
\subsection{Forward Euler Method: Liu et. al. (2023)}\label{apx: apx1-1}
\cite{Liu_An_Fang_Wang_Low_Jordan_2023} solved the history-state quantum encoding problem (i.e., problem 2). They use the forward Euler method to solve the Carleman ODE arising from a reaction-diffusion equation. 

The Carleman ODE is a first-order homogeneous ODE. 
\begin{equation}
  \frac{\d{  \yy}}{\d{t}} = A  \yy, \qquad
   \yy(0) =  \yy_{\mathrm{in}}
\label{eq:LODE4}
\end{equation}

The time interval $[0, T]$ is uniformly discretized in $m = T/h_t$ sub-intervals to solve using the forward Euler method.

\makebox[0pt][l]{%
\begin{minipage}{\textwidth}
\centering
    \includegraphics[width=.5\textwidth]{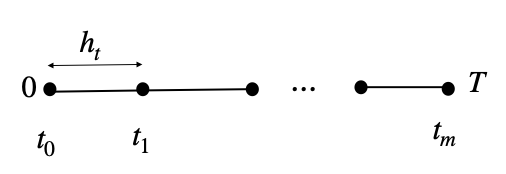}
 \captionof{figure}{Uniform time steps between $[0,\ T]$}
 \label{fig:time-steps}
\end{minipage}
}

The recurrence relation is given by (assume $h:=h_t$)

\begin{equation}
y^{k+1} = [I+Ah] y^k 
\label{eq:forward}
\end{equation}

where component $ y^k \in \R^{\Delta}$ approximates the value of $\yy(t=kh_t)$ for each $k \in \rangez{m+1} \coloneqq \{0,1,\ldots,m\}$. The base case $\yy^0 = \yy(t=0):= \yy_{\mathrm{in}}$. 
This gives an ($(m+1)\Delta) \times ((m+1) \Delta$) sized linear system

\renewcommand{\arraystretch}{1.4}
\begin{equation}\label{eq:linear_system}
  \begin{pmatrix}
    I &  &  &  &  \\
    -(I\!+\!Ah) & I &  &  &  \\
     & \ddots & \ddots &  &  \\
     &  & -(I\!+\!Ah) & I &  \\
     &  &  & -(I\!+\!Ah) & I \\
  \end{pmatrix}
  \begin{pmatrix}
    y^0 \\
    y^1 \\
    \vdots \\
   y^{m-1} \\
    y^m \\
  \end{pmatrix}
=
  \begin{pmatrix}
    y_{\mathrm{in}} \\
    0 \\
    \vdots \\
    0 \\
    0 \\
  \end{pmatrix}.
\end{equation}
\renewcommand{\arraystretch}{1}

\textbf{Quantum encoding of the linear system:} This is encoded in quantum form as
\begin{equation}
L|\YY\rangle=|B\rangle, 
\label{eq:linear_system}
\end{equation}
where
\begin{equation}
L = \sum_{k=0}^{m+1}|k\rangle\langle k|\otimes I-\sum_{k=1}^{m+1}|k\rangle\langle k-1|\otimes (I+Ah).
\label{eq:matrixL}
\end{equation}

\begin{equation}
\ket{\YY} = \frac{1}{\||\YY\rangle\|}\sum_{k=0}^{m}y^k|k\rangle = \frac{1}{\|\YY\|}\sum_{k=0}^{m}\sum_{j=1}^Ny_j^k|j\rangle|k\rangle
\end{equation}
where the normalization factor satisfies $\||\yy\rangle\|^2 = \sum_{k=0}^{m}\|y^k\|^2 = \sum_{k=0}^{m}\sum_{j=1}^N\|y_j^k\|^2$. 
Also 
\begin{equation}
|B\rangle = \yy_{\mathrm{in}}\ket{k=0}
\end{equation}

The quantum (encoded) linear equation \eq{linear_system} encodes the recurrence relation \eq{forward}. 

The state encoding can be explicitly written as: 
\begin{align}\label{eq:detail_y_jk}
    \YY = [y^0,\ y^1,\cdots, y^m]^T\\
    where,\ y^k = [y^k_1,\ y^k_2,\cdots, y^k_N]^T\ \ \forall k\in \{0, ..., m\}\\
     and \ y^k_j = [u(t= kh)]^{\otimes j}
\end{align}

\textbf{The QLSA solver:} This is solved using the QLSA mentioned in \cite{childs2017quantum}, whose output is a quantum state vector encoding the solution $\YY$ of the linear system \eq{linear_system}.  

\textbf{Conditional measurement:} The goal is to produce the history state encoding
\begin{equation}
    |u_{\mathrm{evo}}\rangle = \sum_{k=0}^{m} u(kh)|k\rangle  = \sum_{k=0}^m y^k_1\ket{j=1}\ket{k}
\end{equation}
It is achieved by conditional measurement on register $\ket{j}$. If the outcome of the register $\ket{j}$ is one, success is declared. Otherwise, the algorithm is repeated. \QEDB

$\mathsf{Reason:}$ \cref{eq:detail_y_jk} implies the measurement outcome of $j=1$ would collapse the quantum state $\yy$ to the desired state
\begin{equation}
    |u_{\mathrm{evo}}\rangle:=\ket{y_1^k} = \ket{u(0)} + \ket{u(h)} + \cdots + \ket{u(mh)} = \sum_{k=0}^{m} u(kh)|k\rangle 
\end{equation}

\textbf{Success probability on conditional measurement:} It is crucial to figure out how probable the event of getting $j=1$ is on conditional measurement. Theorem 6 (\cite{Liu_An_Fang_Wang_Low_Jordan_2023}) provides a lower bound on the probability of measuring the state corresponding to $j=1$ as follows. 
\begin{theorem}\label{theorem:measure_prob}
    Then the probability of measuring a quantum state $\ket{u_{\mathrm{evo}}}=\sum_{k=0}^m |y^k_1\rangle$ satisfies
\begin{equation}
P_{\mathrm{(j=1)}} \ge \frac{2G^2}{16\max_{t\in[0,T]}\|\yy(t)\|^2 + G^2}.
\end{equation}
\end{theorem}
Where 
\begin{equation}
G \coloneqq \sqrt{\frac{\sum_{k=0}^{m}\|\yy_1(kh)\|^2}{m+1}},
\label{eq:g}
\end{equation}
and due to equation 4.54 in \cite{Liu_An_Fang_Wang_Low_Jordan_2023}
\begin{equation}
    \max_{t\in[0,T]}\|\yy(t)\|^2 \leq ||u_{\mathrm{in}}||^{4N}
\end{equation}
\QEDB

\textbf{Amplitude implication:} This is required to improve the success of measuring the desired state after conditional measurement. The last theorem implies 
$$\Or(\sqrt{1/P_{\mathrm{(j=1)}}})= O\left(\frac{u_{\mathrm{in}}^{2N}}{G}\right)$$ 
Thus, this many rounds of amplitude amplification subroutine on $\ket{\yy}$ is sufficient to boost the success of measuring $j=1$ to nearly 100 percent. \QEDB

\textbf{Claim:} The gate complexity of the whole algorithm is given by
\begin{equation}
    \mathsf{O(C_{amp}\cdot C_{QLSA})}
\end{equation}
\textbf{Reason:} We can see QLSA as a subroutine to produce the quantum state $\ket{\yy}$. The amplitude amplification has improved the chance of measuring the desired output $\ket{u_{\mathrm{evo}}}$ (subspace) on the conditional measurement on $\ket{\yy}$ (space).  \QEDB

\textbf{Computational cost:} 
\begin{itemize}
    \item The gate complexity of amplitude amplification is known to be optimal.
    \item Different variants of QLSA are known to have different gate complexities.
\end{itemize}

\cite{Liu_An_Fang_Wang_Low_Jordan_2023} use theorem 5 from \cite{childs2017quantum} whose gate complexity is 
\begin{theorem}[\cite{childs2017quantum}] 
Let $M|x\rangle = |b\rangle$ be an $s$-sparse quantum linear system problem with matrix size $n$ and condition number $\kappa$. Assume oracles for matrix $M$ and  $b$  are provided. Then, there exists a quantum algorithm that produces a state $\varepsilon$ approximate to the analytic solution state $|x\rangle$ with gate complexity 
\begin{equation}
    \mathsf{O(s\cdot \kappa\cdot polylog( s\kappa / \varepsilon )\Big(log(n)+polylog(\kappa/\varepsilon))\Big)}
\end{equation}
\label{eq:qlsa_cks}
\end{theorem}
Since there is an upper bound on $P_{(j=1)}$ (\cref{theorem:measure_prob}) and condition number $\kappa$ (equation 4.29 and 4.51 \cite{Liu_An_Fang_Wang_Low_Jordan_2023}) in terms of system parameters. The final complexity of the problem is given in the following theorem.

\begin{theorem}[Theorem 4.1 in \cite{Liu_An_Fang_Wang_Low_Jordan_2023}]
 For given quadratic ODE problem-1 with parameter $R<1$, 
 there exists a quantum algorithm that produces a state that approximates $u_{\mathbf{evo}} $ succeeding with probability $\Omega(1)$ with the query complexity (to the oracles $O_{F_1}, O_{F_2}$, and $O_{u_{\mathrm{in}}}$) 
\begin{equation}
    \begin{aligned}
    \mathsf{\frac{1}{G^2\epsilon}sT^2D^2n^4N^3\|u_{\mathrm{in}}\|^{2N} \biggl(polylog\frac{aDnNsT}{G\epsilon}\biggr)}.
    \end{aligned}
\end{equation}
The gate complexity is larger than its query complexity by logarithmic factors. \QEDB
\label{thm:main}
\end{theorem}

Where, notation 
\begin{itemize}
    \item $G$ is normalization factor (in equation \eq{g}),
    \item $s$ is the sparsity of $F_1$ matrix,
    \item $T$ is the simulation time,
    \item $n$ is the number of spatial discretization points,
    \item $N$ is the Carleman truncation order
    \item $||u_{\mathrm{in}}||$ is the norm of the initial state ($\mathsf{u(t=0)}$),
    \item $a$ and $b$ are parameters of the Fischer-KPP equation
\end{itemize}

\textbf{Conclusion:} The gate complexity scales polynomially in several parameters. It leaves the question can the dependency on $\epsilon$ be improved from polynomial in ($\frac{1}{\varepsilon}$) to polynomial in $\mathsf{log(\frac{1}{\varepsilon})}$. Also, the parameter $N$ appears in the power of $||u_{in}||$, thus an exponential scaling. These shortcomings where improved by $\cite{costa2025further}$ to $\mathsf{polylog(\frac{1}{\varepsilon})}$. It also improves the dependency on $N$ from exponential to polynomial. In the next section, we see these details.
\subsection{Truncated Taylor series method: Costa et. al. (2023)}\label{apx: apx1-2}
The runtime of the last algorithm is improved by Costa et al. (\cite{costa2025further}). They introduced two key changes as 
\begin{enumerate}
    \item Rescaling the quadratic ODE problem before linearization (say by a factor $1/\gamma$)
    \item Solving the Carleman ODE by the truncated Taylor series method
\end{enumerate}
Later, the algorithm's output is multiplied by the inverse of the scaling factor $\gamma$ to retrieve the desired solution. 

It should be noted that the paper \cite{costa2025further} solves the variant \ref{prob:prob1} of the problem that demands a quantum vector $\ket{u(T)}$ rather than the history state 
$$\ket{u_{\mathrm{evo}}}= \sum_{k=0}^{m} u(kh)|k\rangle.$$ As we will discern later, the key difference is in how conditional measurement is done on the output of the QLSA subroutine ( i.e., $\ket{\yy}$).

\begin{definition}[Rescaled Quadratic ODE problem]
     Consider a nonlinear ODE system of the form $\diff{u}{t} = F_1 u + F_2 \uu^{\otimes 2}$ as in \cref{eq:sucE}. Then, using a variable transformation in the form of a rescaling $\widetilde{u}= u/\gamma$ with $\gamma>0$, we obtain another system in the rescaled variable 
    \begin{equation}
    \label{eq:resc_ODE}
        \diff{\widetilde{u}}{t} = \widetilde{F}_1 \widetilde{u} + \widetilde{F}_2 \widetilde{u}^{\otimes 2} , \quad 
    \end{equation}
    with $\widetilde{F}_1 = F_1$ and $\widetilde{F}_2 = \gamma F_2$. \QEDB
    \label{defn:rescaled-carleman}
\end{definition}

$\mathsf{Good\ choice\ for\ the\ scaling\ factor:}$ Equation 55 of \cite{Costa_Schleich_Morales_Berry_2023} prescribes taking the scaling factor as per
\begin{equation}
    \gamma \leq \frac{||u_{\mathrm{in}}||}{R}.
\end{equation}

$\mathsf{Scaled\ Carleman\ ODE:}$ The scaling factor slightly alters the Carleman ODE problem. We get 
\begin{align}
    \widetilde{A}_j^{j} = A_j^{j}\ and\ \widetilde{A}^{(j)}_{j+1} = \gamma A^{j}_{j+1} \\
    \widetilde{\yy} = [\widetilde{u},\ \widetilde{u}^{\otimes 2}, \cdots,\ \widetilde{u}^{\otimes N}]
\end{align}
As a result, we can write the rescaled Carleman linearization as
\begin{equation}
\dt{\widetilde{\yy}} = \widetilde{{A}}\widetilde{\mathbf{y}}, 
\end{equation}
where 
\begin{equation}\label{eq:resccarlmatrix}
\widetilde{A}=
\begin{bmatrix}
A_1^{1} & \gamma A^{1}_{2}          & \cdots & 0        & & 0             & \cdots       & 0          \\
0         & A_2^{2}  & \gamma A^{2 }_{3} & 0        & 0          & 0  & 0            & \vdots     \\
\vdots    & 0          & \ddots &          &            & 0             & \ddots       & 0          \\
          &            & \ddots & \ddots   &            &               & \ddots       & 0  \\
          &            &        &          & \ddots     &               &              & 0          \\
          &            &        &          & \ddots     & \ddots        &              & \vdots     \\
\vdots    &            &        &          &            & 0             & A_{N-1}^{N-1 }& \gamma A_{N-1}^{N}          \\
0         & 0          & \cdots &          &            & \cdots        & 0            & A_N^{N }  \\
\end{bmatrix} .
\end{equation}

\textbf{Note:} We drop the tilde (symbol) above matrix $A$ and vector $\yy$ for convenience. 

\textbf{The QLSA solver:} This homogeneous and time-independent ODE is solved using the Taylor series method as described in reference \cite{berry_costa22}. This output the quantum vector $\ket{\widetilde{\yy}}$.

\textbf{Conditional measurement:} Since the goal of the problem 2 is to output $\ket{u(T)}$. This is recovered by conditional measurement on 
\begin{equation}
    \mathsf{\widetilde{\yy} = \sum_{j=1}^N\widetilde{y_j}\ket{j}\ for \ j\in \{1, ..., N\}.}
\end{equation}
You measure the register $\ket{j}$ and declare success only if you measure $j=1$. Otherwise, the algorithm is repeated until successful.

\textbf{Probability of the success on conditional measurement:} \cite{Costa_Schleich_Morales_Berry_2023} gives an estimate of how probable this event is in the following theorem
\begin{theorem}[Lemma 2 in \cite{Costa_Schleich_Morales_Berry_2023}] 
    For the scaled Carleman ODE with the truncation number $N$ and parameter $R<1$, the probability of getting $j=1$ is given by 
    $P(\widetilde{y_1}) \geq\frac{1}{N}$.
\end{theorem}
\textbf{Note 1:} We have skipped the internal details of the ODE solver \cite{berry_costa22}. In fact, they approximate the solution using the $k$-th order Taylor series, which provides a recurrence relation similar to the forward Euler method. Then, they use a more optimized version of $\mathsf{QLSA}$ solver \cite{Costa_An_Sanders_Su_Babbush_Berry_2022}, instead of \cite{childs2017quantum}.

\textbf{Note :} The gate complexity of the solver \cite{Costa_An_Sanders_Su_Babbush_Berry_2022} depends on the gate complexity of the block encoding of the matrices. They have computed the cost of such an operation using input parameters.

\begin{theorem}[Lemma 5 in \cite{Costa_Schleich_Morales_Berry_2023}] For given quadratic ODE problem-2 with parameter $R<1$, there exists a quantum algorithm that produces a state that approximates $u(T)$ succeeding with probability $\Omega(1)$ with the query complexity to the oracles $O_{F_1}$ and $O_{F_2}$
\begin{equation}
    \mathsf{\order{
    \frac{1}{\sqrt{1-R^{2}}}\frac{\uin}{\norm{\uu(T)}} (|a| + (4\pi^2/3)Dn^2) T N\log({\frac{N}{\varepsilon}}) \log(\frac{ N (|a| + (4\pi^2/3)Dn^2) T}{\varepsilon})},}
\end{equation}
and query complexity to oracles for $\uu_{\mathrm{in}}$ (say, $O_{u_{\mathrm{in}}}$), 
\begin{equation}
\label{eq:nonODE_comp}
    \mathsf{\order{
    \frac{1}{\sqrt{1-R^{2}}}\frac{\uin}{\norm{\uu(T)}} (|a| + (4\pi^2/3)Dn^2) T N^{2}\log({\frac{N}{\varepsilon}}) },}
\end{equation}
and an \textbf{additional} gate of order 
\begin{equation}
    \mathsf{\order{
    \frac{1}{\sqrt{1-R^{2}}}\frac{\uin}{\norm{\uu(T)}} (|a| + (4\pi^2/3)Dn^2) T N^{2} \log({\frac{N}{\varepsilon}}) \log(\frac{ N (|a| + (4\pi^2/3)Dn^2) T} {\varepsilon})^2\log n},}
\end{equation}
is required. The Carleman truncation order scales as 
\begin{equation}
    \mathsf{N = \order{\frac{\log\left(1/\varepsilon\right)}{\log\left(1/R\right)}}}.
\end{equation}
With the assumption of $\lambda_{F_2}/\|F_2\| = \mathcal{O}(\lambda_{F_1}/\|F_1\|)$ for the block encoding of $F_1$ and $F_2$.
\label{lem:cos23}
\end{theorem}
Where, notation 
\begin{itemize}
    \item $D$ is the diffusion coefficient,
    \item $a$ is the coefficient of $u(x,t)$ in Fisher-KPP equation,
    \item $R$ is non-linearity parameter (see \eq{A1}),
    \item $T$ is the simulation time,
    \item $n$ is the number of spatial discretization points,
    \item $N$ is the Carleman truncation order,
    \item $||u_{\mathrm{in}}||$ is norm of initial solution (at t=0), and
    \item $\lambda_{F_{1}}$ is $\lambda$-value for the block encoding of matrix $F_1$ (refer to QLSA solver in \cite{Costa_An_Sanders_Su_Babbush_Berry_2022}).
\end{itemize}

\textbf{Conclusion: } Overall, \cite{costa2025further} have improved the $\varepsilon$ dependency from $\mathsf{poly(1/\varepsilon)}$ to $\mathsf{polylog(1/\varepsilon)}$. Also, the term containing the parameter $N$ in the exponent has been removed due to the 'rescaling trick'.

\subsection{Error Analysis and Numerical Simulation}\label{apx: simulation}
In the following subsection, we discuss the stability of the solution resulting from the introduction of various approximation techniques, including Carleman truncation, central difference discretization, and numerical integration schemes (Euler, Taylor, and Chebyshev). We give our MATLAB simulation results for the Fisher-KPP equation with the Dirichlet boundary condition.
\subsubsection{Carleman truncation error}
The error due to finite truncation of Carleman ODE, originating from the reaction-diffusion equation, was first studied by \cite{Liu_An_Fang_Wang_Low_Jordan_2023}. For $R<1$, they proved that the truncation error exponentially converges as the truncation number increases\footnote{Assume the vector norm $||.||$ is the $l_2$ norm. }.

\begin{theorem}[Theorem 3.2 in \cite{Liu_An_Fang_Wang_Low_Jordan_2023}] 
    Let the eigenvalues of the matrix $F_1$ in the quadratic ODE be all real and negative, and $\lambda_1$ be the largest eigenvalue. Then for any $j \in \range{N}$, the truncation error $\eta_j(t) \coloneqq u^{\otimes j}(t)- y_j(t)$ satisfies
    \begin{equation}
        \|\eta_j(t)\| \le \|u_{\mathrm{in}}\|^j R^{N+1-j}(1-e^{\lambda_1 t}).
    \label{eq:bound_general}
    \end{equation}
    For $j=1$, we have the tighter bound
    \begin{equation}
        \|\eta_1(t)\| \le
        \|u_{\mathrm{in}}\| R^N\big(1-e^{\lambda_1 t}\big).
    \label{eq:bound_1}
    \end{equation}
    \label{cor:error_liu23}
\end{theorem}

Later on \cite{Costa_Schleich_Morales_Berry_2023} improved the error bound as follow

\begin{theorem}[Lemma 4 in \cite{Costa_Schleich_Morales_Berry_2023}]
    In the same context as theorem \cref{cor:error_liu23}, the truncation error is given as
    \begin{equation}
        \|\eta_j(t)\| \le \|u_{\mathrm{in}}\|^j R^{N+1-j}f_N(\lambda_1 t).
    \label{eq:bound_general}
    \end{equation}
    For $j=1$, we have the tighter bound
    \begin{equation}
        \|\eta_1(t)\| \le
        \|u_{\mathrm{in}}\| R^Nf_N\big(\lambda_1 t\big).
    \label{eq:bound_1}
    \end{equation}
    Where $f_N(\lambda_1 t)\leq (1-e^{\lambda_1 t})$. 
    \label{cor:error_liu23}
\end{theorem}
The above theorems are characterized by the parameter $R$, which depends on the highest eigenvalue of the matrix $F_1$. Recently, a new regime for convergence has been explored by \cite{wu2025quantum}. It is quantified in terms of a No-resonance parameter that depends on a slightly different property of the eigenvalues of the matrix $F_1$. We have discussed it in \ref{sec:diagonalization} when we mentioned the No-resonance condition.

\makebox[0pt][l]{%
\begin{minipage}{\textwidth}
\centering
    \includegraphics[width=.60\textwidth]{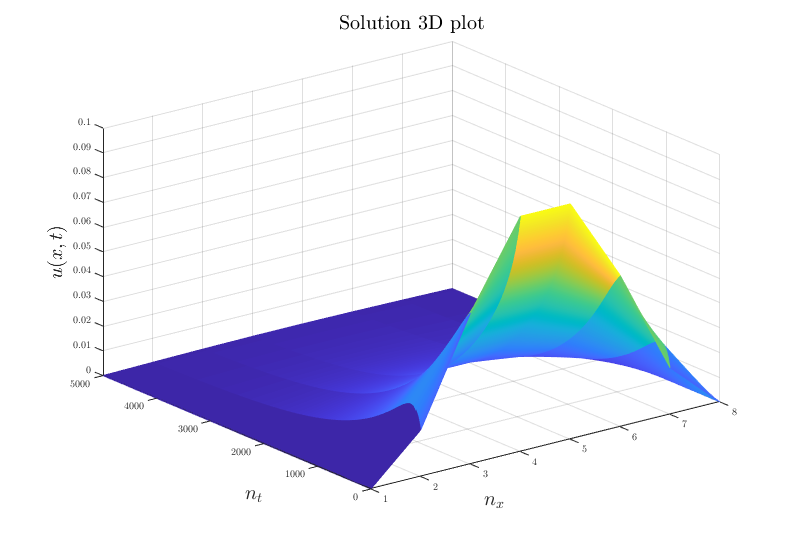}
 \captionof{figure}{(Euler method) Solution for reaction diffusion equation for $D =0.2,\ a=0.4\ and\ b=-1$. The initial distribution $u(x,t=0)= 0.1sin^2(\pi x)$. Position domain $x\in [0,1]$ is discretized into 8 points while time $t\in[0, 3]$ is discretized into 5000 points}. 
 \label{fig:fig_euler-1}
\end{minipage}
}

\makebox[0pt][l]{%
\begin{minipage}{\textwidth}
\centering
    \includegraphics[width=.80\textwidth]{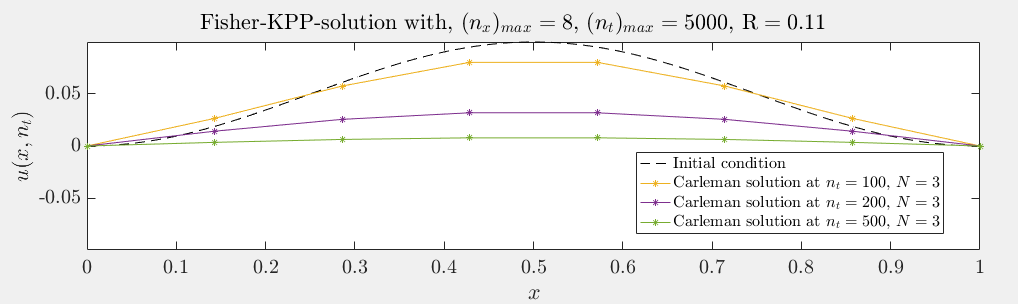}
 \captionof{figure}{(Euler method) Time snapshots at three different times reveal how the solution decays in magnitude as time progresses. Position domain $x\in [0,1]$ is discretized into 8 points while time $t\in[0, 3]$ is discretized into 5000 points}
 \label{fig:fig_euler0}
\end{minipage}
} 

\subsubsection{Error due to ODE solver: Euler \& Taylor method }
This type of error is well-studied in numerical methods for ODE. For the Euler method, the step size is the key determining parameter. 
\begin{theorem}[Lemma 4.3 in \cite{Liu_An_Fang_Wang_Low_Jordan_2023}]
    In the context of  problem 1 with parameter $R<1$, if we select the time step as 
    \begin{equation}
        h\leq\frac{1}{N^2||F_1||}.
    \end{equation}
    Then the error due to the Euler method is bounded as
    \begin{equation}
        \varepsilon_{\mathrm{euler}} \leq N^{2}Th(||F_1||+||F_2||)^2 (max_{t\in [0, T]}||\yy(t)||)
    \end{equation}
\end{theorem}
For the Fisher-KPP equation, the values of these matrix norms are as follows
\begin{align}
    ||F_1|| = 4D(n+1)^2 +a \\
     ||F_2|| = b 
\end{align}

For the $K$-th order truncated Taylor series method, the time propagation error is given as  
\begin{equation}
    \varepsilon_{Taylor} \in \left(\frac{(||A||h)^{K+1}}{(K+1)!}||\yy(0)||\right)
\end{equation}
The spectral norm of the Carleman matrix is known to be bounded by 
\begin{equation}
    ||A|| \leq  N(||F_1|| + ||F_2||)
\end{equation}
It is combined to get the truncation order $K$ as a function of $\varepsilon$ and system parameters.

\makebox[0pt][l]{%
\begin{minipage}{\textwidth}
\centering
    \includegraphics[width=.65\textwidth]{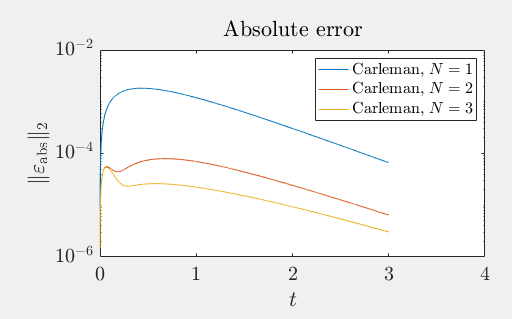}
 \captionof{figure}{(Euler method: absolute error vs time) As time progresses, the error eventually starts decreasing. Time $t\in[0, 3]$ is discretized into 5000 points.}
 \label{fig:fig_euler1}
\end{minipage}
}

\makebox[0pt][l]{%
\begin{minipage}{\textwidth}
\centering
    \includegraphics[width=.65\textwidth]{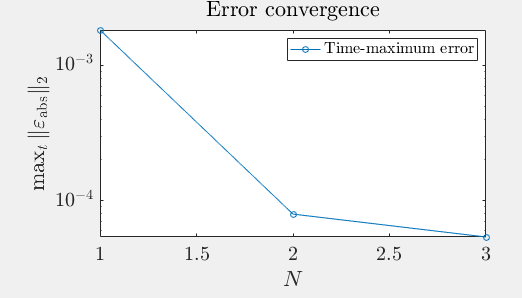}
 \captionof{figure}{(Euler method: time-maximum error vs. Carleman truncation order) As expected, the error decreases as the truncation order increases.}
 \label{fig:fig_euler2}
\end{minipage}
}

\subsubsection{Error due to the spatial discretization}
\cite{Costa_Schleich_Morales_Berry_2023} estimated the error due to uniform finite difference discretization.

\begin{theorem}
    Using a finite difference discretization with $3$ stencil points in one dimension, the solution of the Fisher-KPP equation at time $T>0$ has an error due to spatial discretization when $a<0$ and $|a|> |b|\uin$ bounded as 
\begin{equation}
\|\varepsilon_{\mathrm{disc}}(T)\| = \order{   n^{-1/2}
\norm{\frac{d^3(u(x,t))}{d^3x}} \cdot \frac{1 - \exp{\left( a + |b|{\uin}\right)t} }{ \left|a + |b|{\uin}\right|}},
\end{equation}
where $n$ is the number of grid points used. 
\end{theorem}

$\mathsf{Proof:}$ It is derived from Lemma 7 in \cite{Costa_Schleich_Morales_Berry_2023} for a $d$-dimensional system using $(2k+1)$ stencils. In our case, we have a one-dimensional problem with a $3$-point stencil. Setting $d=1$ and $k=1$ yields the above theorem. \QEDB

\textbf{Remark:} The results on Carleman truncation error and finite difference discretization are applicable in our case, too. We refer to these results in the next section \ref{sec: cheby-simulation} while discussing the sources of error in our algorithms.

\makebox[0pt][l]{%
\begin{minipage}{\textwidth}
\centering
    \includegraphics[width=.65\textwidth]{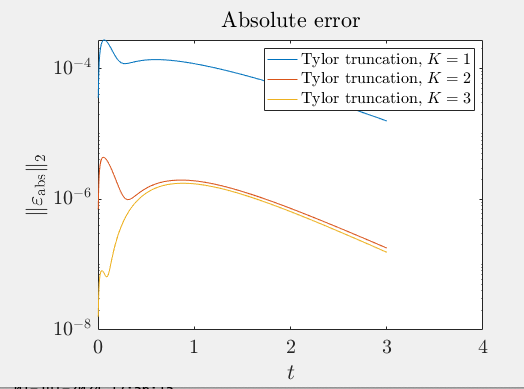}
 \captionof{figure}{(Taylor method: absolute error vs time graph) They corresponds to three different orders of Taylor truncation.  We have fixed the Carleman truncation order to $N=3$. Time step size: the time domain $t$ $\in$ [0, 3] is uniformly divided into 2000 parts for numerical simulation.}
 \label{fig:fig_taylor1}
\end{minipage}
}
\subsection{Numerical Simulation Results} \label{sec: cheby-simulation}
\subsubsection{Numerical Setup} We are solving the Fisher-KPP equation 
\begin{equation}
    \pdv{u(x,t)}{t} = 0.2\Delta u(x,t) + 0.4 u(x,t) - u^2(x,t)
\end{equation}
with initial distribution 
\begin{equation}
    \mathsf{u(x,t=0) = 0.1 sin^2(\pi x)}
\end{equation}
\textbf{Proxy for the analytic solution:} Errors are defined with respect to the analytic solution. We use MATLAB Runge-Kutta 45 ODE solver (MATLAB:ode45) to solve the quadratic ODE problem. We take this solution as a proxy for the analytic solution. It is one of the best ODE solvers available in MATLAB.

Then, the same quadratic ODE is solved using either 
\begin{itemize}
    \item Carleman linearization \& Forward Euler method (fig \ref{fig:fig_euler-1} to \ref{fig:fig_euler2})
    \item Carleman linearization \& Taylor series method (fig: \ref{fig:fig_taylor1} to \ref{fig:fig_taylor2})
    \item Carleman linearization \& Matrix Exponentiation method (fig: \ref{fig:sim-cheby1} to \ref{fig:cheb_error})
\end{itemize}

\textbf{Methodology:} It is a classical numerical result. The quantum algorithm is designed to solve the Euler, Taylor, or Chebyshev method by solving the corresponding linear system of equations. Rather, we solved them iteratively as per the Euler, Taylor, and Chebyshev methods without embedding them into a linear system. The reason is that we are interested in assessing the stability of the solution after combining the Carleman linearization with numerical ODE solvers (the Euler, Taylor, or Chebyshev method). For this purpose, the methodology would yield the desirable result.

\makebox[0pt][l]{%
\begin{minipage}{\textwidth}
\centering
    \includegraphics[width=.65\textwidth]{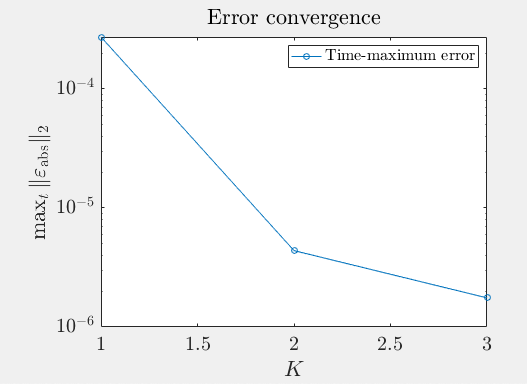}
 \captionof{figure}{(Taylor method: Error vs truncation order) As the Taylor truncation order increases, the time-maximum error decreases.}
 \label{fig:fig_taylor2}
\end{minipage}
}


Now we do numerical simulation for both the Chebyshev-based algorithms.
For a baseline for comparison, we solve the quadratic ODE by the inbuilt solver in MATLAB\footnote{See more details at the official documentation site \href{https://in.mathworks.com/help/matlab/ref/ode45.html}{Matlab:ode45}.}, which gives the following $3$-$D$ graph for the solution $\uu(x,t)$.

\makebox[0pt][l]{%
\begin{minipage}{\textwidth}
\centering
    \includegraphics[width=.60\textwidth]{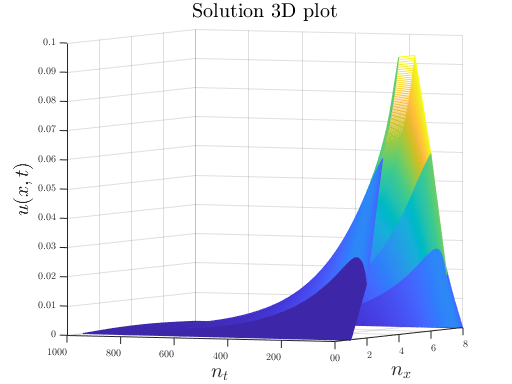}
 \captionof{figure}{(Runge-Kutta 45 Method) Solution for reaction diffusion equation for $D =0.2,\ a=0.4\ and\ b=-1$. The initial distribution $u(x,t=0)= 0.1sin^2(\pi x)$. Position domain $x\in [0,1]$ is discretized into 8 points, while time $t\in[0, 3]$ is discretized into 1000 points.} 
 \label{fig:fig1}
\end{minipage}
}

Solving the Fisher-KPP using the Carleman linearization and Chebyshev series gives the following $3D$ graph for the solution $\uu(x, t)$.

{%
\begin{minipage}{\textwidth}
\centering
    \includegraphics[width=.60\textwidth]{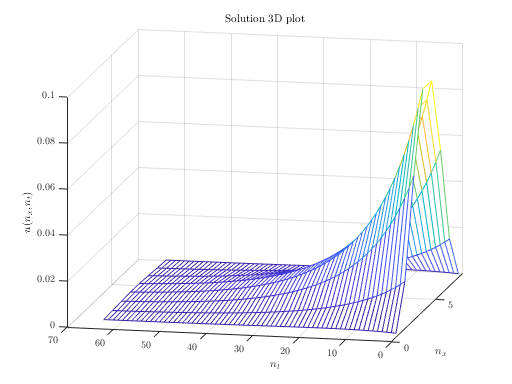}
 \captionof{figure}{(Chebyshev-series method) Solution for reaction diffusion equation for $D =0.2,\ a=0.4\ and\ b=-1$. The initial distribution $\mathsf{u(x,t=0)= 0.1sin^2(\pi x)}$. Position domain $x\in [0,1]$ is discretized into 8 points. To achieve the desired resolution, the function is approximated at 64 points in the time domain $t\in[0,3]$. [ See the \href{https://github.com/108mk/Quantum_Tech_Project_M_Tech_IISc/tree/f4a6f80e24864f84c451d418116c44edb6a8972d/codes/Chebyshev-codes}{GitHub code} for the simulation.]}
 \label{fig:sim-cheby1}
\end{minipage}
}

{%
\begin{minipage}{\textwidth}
\centering
    \includegraphics[width=.70\textwidth]{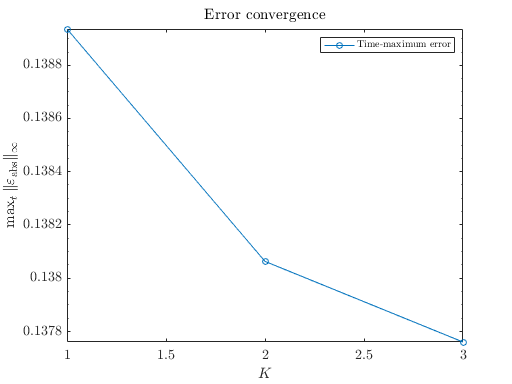}
 \captionof{figure}{(Absolute Error vs Chebyshev truncation order $K$) The Carleman truncation is fixed at $N=3$ while the Chebyshev truncation order is varied from $K=1$ to $3$.}   
 \label{fig:cheb_error}
\end{minipage}

\textbf{Some limitations of the simulation technique:} As it is clear from the algorithm pipeline (\cref{fig:summary_fig}), several approximation techniques have been employed to obtain the final solution. It includes errors due to finite difference discretization, Carleman truncation, and the Chebyshev method used to solve the ODE. An absolute error graph (\cref{fig:cheb_error}) wouldn't differentiate between them. A separate controlled error analysis is required to distinguish between individual errors.

\section{Appendix II: Supplementary Technical Results}\label{apx: apx2}
\subsection{Properties of matrix $F_1$ and $F_2$}
\begin{lemma}\label{lem:lemma1}
The following two properties hold for the matrix $F_1$ 
    \begin{enumerate}
        \item It is a real symmetric matrix, thus Hermitian.
        \item It has real and negative eigenvalues only if parameters $D$, $a$, and $n$ are related by
    \end{enumerate}
\begin{equation}
    \mathsf{4D(n+1)^2sin^2\Big(\frac{j\pi}{2(n+1)}\Big)\geq a}
\end{equation}
\end{lemma}
$\mathsf{Proof}$: The eigenvalues of $L_h$ are given by
\begin{equation}
        \mathsf{\lambda_j(L_h) = -4(n+1)^2sin^2\Big(\frac{j\pi}{2(n+1)}\Big);\  where\ j\in \{1, ..., n\}}
\end{equation}
Since $F_1 = DL_h+aI$, its eigenvalues are given by 
\begin{equation}
        \mathsf{\lambda_j(F_1) = -4D(n+1)^2sin^2\Big(\frac{j\pi}{2(n+1)}\Big)+a;\  where\ j\in \{1, ..., n\}}
\end{equation}
Thus $\lambda_j(F_1)\leq 0$ if 
\begin{align}
        \mathsf{4D(n+1)^2sin^2\Big(\frac{j\pi}{2(n+1)}\Big)\geq a}
\end{align}
Usually, the diffusion coefficient $D\geq 0$. Hence, $a<0$ makes the above inequality true unconditionally. But specific values of $a>0$ are also allowed, as given by the above equation.  \QEDB

Another interesting property of matrices $F_1$ and $F_2$ is that the sparsity parameter (say, $s$) is independent of their size, or $s=O(1)$. 
\begin{lemma}\label{eq:sparse}
    Matrix $F_1$ has sparsity parameter $s =3$, independent of its size $n$. While for $F_2$, sparsity $s=1$.
\end{lemma}
$\mathsf{Proof:}$ The definition of the sparsity parameter $s$ is the maximum number of non-zero entries along rows or columns.  For $F_1$, $s=3$ due to the construction of the discrete Laplacian matrix $L_h$. For $F_2$, it is due to the construction prescribed by its definition. \QEDB

Matrix sparsity plays a crucial role in designing efficient quantum algorithms. We use these results in the later parts.

\subsection{Properties of truncated Carleman matrices}
There are a couple of important lemmas on the sparsity and size of the Carleman matrix that we will use later.
\begin{lemma}\label{lem:sparsity_carleman}
    The $N$-th order truncated Carleman matrix $A$ is an $(3N)$-sparse matrix, where $s =3$ is the sparsity of $F_1$ and $F_2$.
\end{lemma}
$\mathsf{Proof}$: Due to the block structure of the matrix $A$, the sparsity depends on the sparsity of the last blocks $A_N^N$ and $A_N^{N-1}$. These block matrices are constructed using the Kronecker product of the identity matrix with $F_1$ (or $F_2$). We have discussed $F_1$ and $F_2$ have sparsity $s=3\in O(1)$. As a consequence, the sparsity of the block matrix $A_N^N$ is $3N \in O(N)$. \QEDB

The dimension of the Carleman matrix \eq{UODE} is crucial for analysing the complexity of the algorithms.
\begin{lemma}\label{lem:size}
The size of the Carleman matrix  depends on the size of the matrix $F_1$ and the truncation order $N$ as
    \begin{equation}
        d := \Delta:=n+n^2+\cdots+n^N=\frac{n^{N+1}-n}{n-1}=O(n^N).
    \end{equation}
\end{lemma}
Proof: Compute the size of each of the block matrices $A_j^j$. Now, use the fact that the sum of their size is equal to the size of the Carleman matrix.  \QEDB

\subsection{On an upper bound on the condition number}\label{apx: apx2-3}
In Chapter 5, the iterative construction of the matrix $V\in \R^{d\times d}$ is by computing a set of linearly independent vectors for the Carleman matrix $A$. Let $V= [e_1,..., e_d]$. We have an explicit form for each $e_j$.
Using Guggenheimer's result \cite{Guggenheimer95} on the condition number, we get

\begin{equation}
    \kappa (V) < \frac{2}{|Det(V)|} \left(\frac{\|V\|_F}{\sqrt{d}}\right)^d.
\end{equation} 
    
Here, the Frobenius norm owes
\begin{equation}
    \|V\|_F = \sum_{i=1}^d ||e_i||_2.
\end{equation}
The determinant of $V$ can be estimated as it is equal to the product of the determinants of the diagonal block matrices \footnote{See this discussion on Math StackExchange \url{https://math.stackexchange.com/q/1184825/474528} }. It is not obvious how to give an upper bound on $\|A\|_F = \sum_{i=1}^d ||e_i||_2$. We leave it as an exercise for the future. \QEDB

\section{Appendix III: The Diagonalization Analysis for Higher Degree Non-linear ODEs}\label{apx: apx3}
Our analysis on diagonalization can readily be generalized to a higher degree non-linear ODE, like
\begin{equation}
    \frac{\d{u}}{\d{t}} = F_Mu^{\otimes M}+F_1u, \qquad
    u(0) = u_{\mathrm{in}}.
\end{equation}

This is because the associated Carleman matrix for this case has the form
\begin{equation}
    A = \begin{pmatrix}
     A_1^1 & 0 & \cdots & 0 & A_M^1 &  &  \\
     & A_2^2 & \ddots &  & \ddots & \ddots &  \\
     &  & \ddots & \ddots &  & \ddots & A_N^{N-M+1} \\
     &  &  & \ddots & \ddots &  & 0 \\
     &  &  &  & \ddots & \ddots & \vdots \\
     &  &  &  &  & A_{N-1}^{N-1} & 0 \\
     &  &  &  &  &  & A_N^N \\
  \end{pmatrix}
\end{equation}

Due to \cref{thm:no_reso_eigs}, its eigenvalue entirely depends on the eigenvalues of the block matrices along the main diagonal. Thus, changing the degree of the polynomial has no impact on the eigenvalues of the new Carleman matrix. It does impact the similarity transformation matrix $V$. Again, the iterative procedure for diagonalization can be described because the matrix remains a block matrix with a bi-diagonal structure.

The problem of diagonalization requires more careful consideration if there are mixed terms in the non-linear ODE, like
\begin{equation}
    \frac{\d{u}}{\d{t}} = \sum_{k=2}^MF_ku^{\otimes k}+F_1u, \qquad
    u(0) = u_{\mathrm{in}}.
\end{equation}
Now, the Carleman matrix is no longer a block bi-diagonal matrix. It appears that the iterative procedure for diagonalization needs to be substantially changed. It can be seen as an open problem related to this work. 
\QEDB

\end{document}